\documentclass[journal,a4paper,final,twocolumn,10pt,twoside]{IEEEtranTCOM}

\usepackage{layout}
\normalsize

\ifCLASSINFOpdf

\else

\fi

\usepackage{mathtools}
\usepackage{amsmath}
\usepackage{amssymb}
\usepackage{amsmath}
\usepackage{cite}
\usepackage{lipsum}
\usepackage{cuted}
\usepackage{multirow}
\usepackage{multicol}
\usepackage{siunitx}
\usepackage{graphicx}
\usepackage{graphicx}
\usepackage{slashbox}
\usepackage{mdwmath}
\DeclareMathOperator{\E}{\mathbb{E}}
\usepackage{eqparbox}
\usepackage{fixltx2e}
\date{}
\begin{document}

\title{Performance Studies of Underwater Wireless Optical Communication Systems with Spatial Diversity: MIMO Scheme} %
\author{Mohammad Vahid Jamali,~\IEEEmembership{Student Member,~IEEE}, Jawad A. Salehi,~\IEEEmembership{Fellow,~IEEE}, and Farhad Akhoundi
\thanks{Part of this paper is supported by Iran National Science Foundation (INSF). The authors are with the Optical Networks Research Laboratory (ONRL), Department of Electrical Engineering, Sharif University of Technology, Tehran, Iran (E-mail: mohammad.v.jamali@gmail.com; jasalehi@sharif.edu; akhoundi@alum.sharif.edu).
This paper was presented in part at the IEEE International Workshop on Wireless Optical Communication (IWOW), Istanbul, Turkey, September 2015.}
}
\maketitle
\begin{abstract}
In this paper, we analytically study the performance of multiple-input multiple-output underwater wireless optical communication (MIMO UWOC) systems with on-off keying (OOK) modulation. To mitigate turbulence-induced fading, which is amongst the major degrading effects of underwater channels on the propagating optical signal, we use spatial diversity over UWOC links. Furthermore, the effects of absorption and scattering are considered in our analysis. We analytically obtain the exact and an upper bound bit error rate (BER) expressions for both optimal and equal gain combining. In order to more effectively calculate the system BER, we apply Gauss-Hermite quadrature formula as well as approximation to the sum of lognormal random variables. We also apply photon-counting method to evaluate the system BER in the presence of shot noise. Our numerical results indicate an excellent match between the exact and upper bound BER curves. Also {a good match} between {the} analytical results and numerical simulations confirms the accuracy of our derived expressions. Moreover, our results show that spatial diversity can considerably improve the system performance, especially for channels with higher turbulence, e.g., a $3\times1$ MISO transmission in a $25$ {m} coastal water link with log-amplitude variance of $0.16$ can introduce $8$ {dB} performance improvement at the BER of $10^{-9}$.
\end{abstract}
\IEEEpeerreviewmaketitle
\section{Introduction}
\IEEEPARstart{U}{nderwater} wireless optical communications (UWOC) {has recently been introduced to meet a number of demands in various underwater applications due to its scalability, reliability, and flexibility. As opposed to its traditional counterpart, namely acoustic communication, optical transmission has higher bandwidth, better security, and lower time latency. This enables UWOC as a powerful alternative for the requirements of high-speed and large-data underwater communications such as imaging, real-time video transmission, high-throughput sensor networks, etc. \cite{tang2014impulse,hanson2008high,arnon2010underwater,kaushalunderwater,lanbo2008prospects}. However, despite all the above advantages, due to {the} severe degrading effects of the UWOC channel, namely absorption, scattering, and turbulence \cite{mobley1994light,korotkova2012light}, currently UWOC is only appropriate for short-range communications (typically less than $100$ {m}) with realistic average {transmitted powers}. Removing this impediment, in order to make the usage of UWOC more widespread, necessitates comprehensive research on the nature of these degrading effects and also on the intelligent transmission/reception methods and system designs.}


{ In past few years, various studies have been carried out, theoretically and experimentally, to characterize {the} absorption and scattering effects of different water types. Mathematical modeling of a UWOC channel and its performance evaluation using radiative transfer theory have been presented in \cite{jaruwatanadilok2008underwater}. A novel non-line-of-sight UWOC network has been proposed in \cite{arnon2009non}, based on the back-reflection of the propagating optical signal at the ocean-air interface. In \cite{farr2010integrated}, a hybrid optical/acoustic communication system has been developed. The beneficial application of error correction codes in improving the reliability and robustness of UWOC systems {has been experimented} in \cite{simpson2010}. Based on experimental results in \cite{mobley1994light} and \cite{petzold1972volume} for absorption and scattering of different water types, the beam spread function and the channel fading-free impulse response were characterized in \cite{cochenour2008characterization} and \cite{tang2014impulse}, respectively. Furthermore, a cellular underwater wireless network based on optical code division multiple access (OCDMA) technique has been proposed in \cite{akhoundi2015cellular}, while the potential applications and challenges of such an underwater network and the beneficial application of serial relaying on its users' performance have very recently been investigated in \cite{akhoundi2016cellular} and \cite{jamali2015ocdma}, respectively.}



{ Unlike acoustic links where multipath reflection induces fading on acoustic signals, in UWOC systems optical turbulence is the major cause of fading on the propagating optical signal through the turbulent seawater \cite{tang2013temporal}. Optical turbulence occurs as a result of random variations of refractive index. These random variations in underwater medium mainly result from fluctuations in temperature and salinity, whereas in atmosphere they result from inhomogeneities in temperature and pressure changes \cite{tang2013temporal,navidpour2007ber}.
  Recently, some useful results {have been reported in the literature} to characterize {the} underwater fading statistics. A precise power spectrum has been derived in \cite{nikishov2000spectrum} for the fluctuations of turbulent seawater refractive index. Based on this power spectrum, {the Rytov method} has been applied in \cite{korotkova2012light,ata2014scintillations} to evaluate the scintillation index of optical plane and spherical {waves propagating in an underwater} turbulent medium.
     Also Tang \textit{et al.} \cite{tang2013temporal} have shown that temporal correlation of irradiance may be introduced {by a moving medium} and they investigated temporal statistics of irradiance in moving ocean with weak turbulence. Furthermore in \cite{gerccekciouglu2014bit}, the on-axis scintillation index of a focused Gaussian beam has been formulated in weak oceanic turbulence, and {by considering a lognormal} distribution for intensity fluctuations, the average BER has been evaluated.} {Moreover, the average BER and scintillation index of multiple-input single-output (MISO) UWOC links have very recently been analyzed in \cite{dong2016ber} and \cite{gokcce2016scintillation}, respectively, when degrading effects of inter-symbol interference (ISI) are neglected. Additionally, experimental studies on the fading statistics of UWOC channels in the presence of air bubbles have quite recently been carried out in \cite{jamali2016statistical}.}
     \begin{figure}
         \centering
         \includegraphics[width=3.6in]{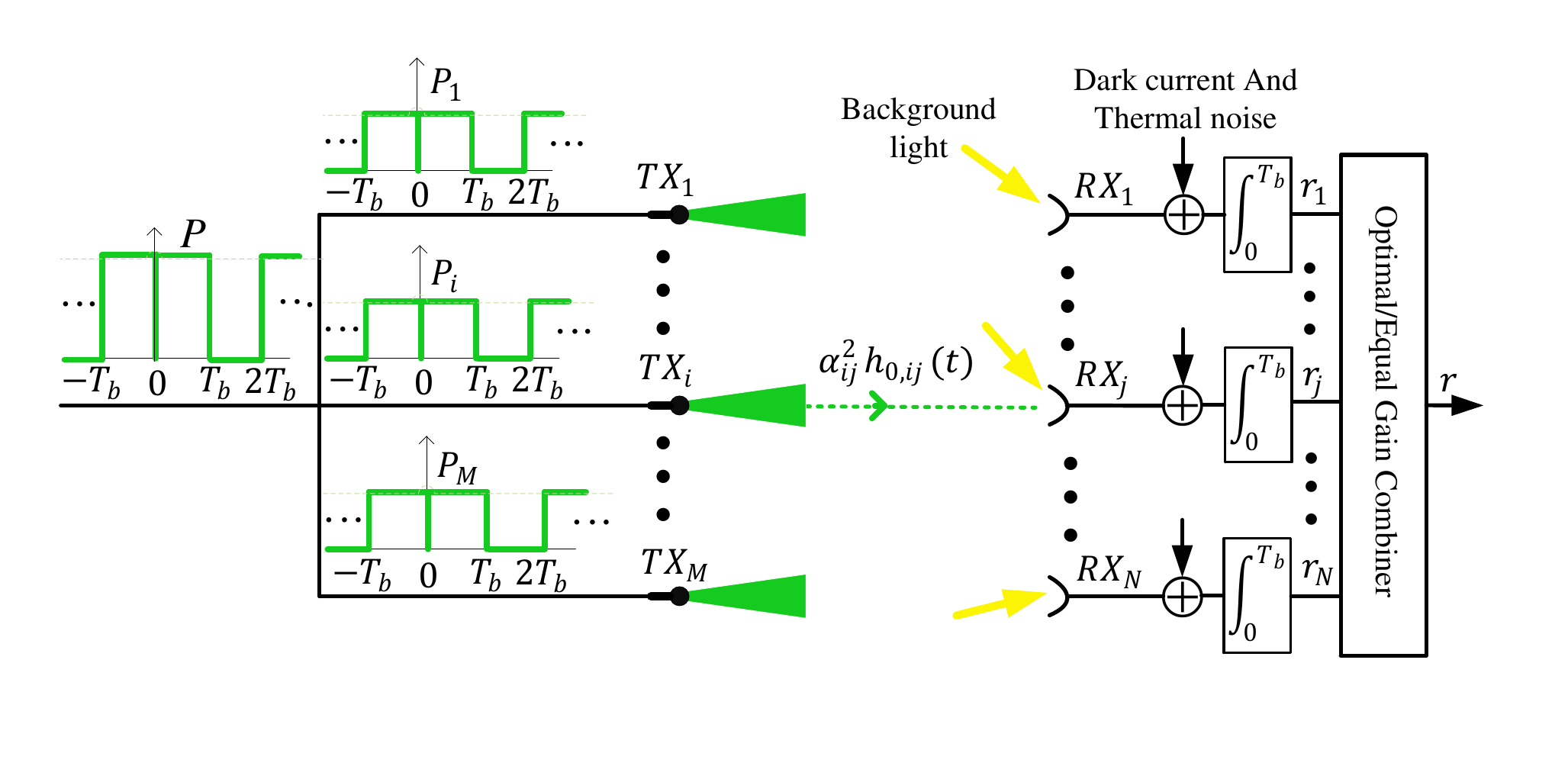}
         \caption{Block diagram of the MIMO UWOC system with OOK modulation.}
         \label{Fig1}
         \end{figure}
   
\indent{{Despite all of the valuable research} have been done on various aspects of UWOC, a comprehensive study on the performance of UWOC systems that takes all degrading effects of the channel into account is missing in the literature. While some {work} only considered absorption and scattering effects and used numerical simulations, without analytical calculations, to estimate the BER of point-to-point UWOC systems \cite{tang2014impulse}, some others considered turbulence effects, but neglected scattering and ISI effects, and used the common conditional BER expressions in {free-space optics (FSO)} \cite{gerccekciouglu2014bit,yi2015underwater}. The research in this paper is inspired by the need to comprehensively evaluate the BER of UWOC systems with respect to all impairing effects of the channel. Moreover, we employ multiple-input multiple-output (MIMO) transmission to investigate how MIMO technique can mitigate turbulence effects and extend the viable communication range.}
 {The technique of spatial diversity}, i.e., exploiting multiple transmitter/receiver apertures (see Fig. \ref{Fig1}), not only compensates for fading effects, but can also effectively decrease the possibility of temporary blockage of the optical beam by obstruction (e.g., {fish}). Another advantage of spatial diversity is in reducing the transmit power density by dividing the total transmitted power by the number of transmitters. In other words, {depending on the wavelength used} there exist some limitations on the maximum allowable safe {transmitted power}. Hence, by employing spatial diversity one can increase the total transmitted power by the number of transmitters and therefore support longer distances, while maintaining the safe transmit power density \cite{navidpour2007ber}. 
 
 {In this paper, we apply maximum-likelihood (ML) detection to analytically obtain the BER expressions for MIMO UWOC systems with repetition coding across the transmitters\footnote{Note that because of the positive nature of incoherent wireless optical communications, a non-destructive addition of different transmitters' intensity signals appears at the receiver; therefore, full channel diversity can be achieved by using simple repetition coding, i.e., sending the same on-off keying (OOK) symbol from all of the transmitters for each bit interval.}, when either an optimal combiner (OC) or an equal gain combiner (EGC) is used. In order to take the absorption and scattering effects into account, we obtain the channel impulse response {using Monte Carlo (MC)} simulations similar to \cite{tang2014impulse,cox2012simulation}. Moreover, to characterize the fading effects, we multiply the above impulse response by a fading coefficient modeled as a lognormal random variable (RV) for weak oceanic turbulence \cite{gerccekciouglu2014bit,yi2015underwater}. Closed-form solutions for the exact and upper bound BER calculation are provided using the Gauss-Hermit quadrature formula as well as approximation to the sum of lognormal RVs. We also apply the photon-counting approach to evaluate the BER of various configurations in the presence of shot noise.}
    In our numerical results we consider both spatially independent and spatially correlated links. 
    
 The rest of the paper is organized as follows. In Section II, we review some necessary theories in the context of our proposed MIMO UWOC system and the channel under consideration in this paper.
In Section III, we analytically obtain both the exact and upper bound BER expressions, when either OC or EGC is used at the receiver side. We also apply Gauss-Hermite quadrature formula to effectively calculate multi-dimensional integrals with multi-dimensional finite series. In order to evaluate the system BER using photon-counting methods, the same steps as Section III are followed in Section IV. Section V presents the numerical results for various system configurations and parameters considering lognormal distribution for fading statistics of the UWOC channel. And Section VI concludes the paper.
 
\section{Channel and System Model}
In this section,
we present the channel model that includes all three impairing effects of the medium, followed by the assumptions and system model that we have introduced in this paper.
\subsection{Absorption and Scattering of UWOC Channels}
 Propagation of optical beam in underwater medium induces interactions between each photon and seawater particles either in the form of absorption or scattering. Absorption is an irremeable process where photons interact with water molecules and other particles and lose their energy thermally. On the other hand, in the scattering process each photon's transmit direction alters, which also can cause energy loss since { fewer photons} will be captured by the receiver aperture. Energy loss due to absorption and scattering can be characterized by absorption coefficient $a(\lambda )$ and scattering coefficient $b(\lambda )$, respectively, where $\lambda$ denotes the wavelength of the propagating light wave. 
{Moreover, total effects of absorption and scattering on the energy loss can be described by extinction coefficient $c\left(\lambda \right)=a\left(\lambda \right)+b(\lambda )$.}
 These coefficients can vary with source wavelength $\lambda $ and water types \cite{tang2014impulse}. It has been shown in \cite{mobley1994light,bohren2008absorption} that absorption and scattering have {their minimum effects} at the wavelength interval $400$ {nm} $<\lambda <530$ {nm}. Hence, UWOC systems apply the blue/green region of the visible light spectrum to actualize data communication.

In \cite{tang2014impulse,cox2012simulation} the channel impulse response has been simulated based on MC method with respect to the absorption and scattering effects. 
 We also simulate the channel impulse response similar to \cite{tang2014impulse,cox2012simulation} relying on MC approach. In this paper, the fading-free impulse response between the $i$th transmitter and the $j$th receiver is denoted by $h_{0,ij}(t)$. As it is elaborated in \cite{tang2014impulse}, when the source beam divergence angle and the link distance increase the channel introduces more {ISI} and loss on the received optical signal. On the other hand, increasing the receiver field of view (FOV) and aperture size increases the channel delay spread while decreasing its loss. { It is {worth mentioning} that although this behavior may be unobservable in clear ocean links, as the water turbidity increases and multiple scattering dominates it becomes more apparent.}
\subsection{Fading Statistics of UWOC Channels}
 In the previous subsection we described how the absorption and scattering effects are characterized. To take turbulence effects into account, we multiply $h_{0,ij}\left(t\right)$ by a multiplicative fading coefficient ${\alpha }^2_{ij}$ \cite{navidpour2007ber,andrews2001laser,zhu2002free,tsiftsis2009optical,karimi2009ber,karimi2011free}, with lognormal distribution for weak oceanic turbulence \cite{gerccekciouglu2014bit,yi2015underwater}. {The assumption of lognormal distribution for UWOC turbulence-induced fading statistics is a reasonable assumption. In fact, UWOC systems typically employ large receiving apertures due to the negligibility of background noise under water which leads to an effectively reduced turbulence. Furthermore, although the attenuation due to pointing error and misalignment for FSO links can be modeled as a RV, for the sake of simplicity and owing to the highly-scattering nature of UWOC channels, the perfect alignment is also assumed in this paper.} To model turbulence-induced fading let $\alpha ={\rm exp}(X)$ be the fading amplitude of the channel with lognormal probability density function (PDF) as \cite{safari2008relay};
 \begin{equation} \label{f(alpha)}
f_{\alpha}({\alpha})=\frac{1}{\alpha \sqrt{2\pi {\sigma }^2_X}}{\rm exp}\left(-\frac{{\left({{\rm ln}  \left(\alpha \right)\ }-{\mu }_X\right)}^2}{2{\sigma }^2_X}\right).
 \end{equation}
Therefore, the fading log-amplitude $X$ has a Gaussian distribution with mean ${\mu }_X$ and variance ${\sigma }^2_X$.
  To ensure that fading neither amplifies nor attenuates the average power, we normalize fading amplitude such that $\E\left[{\alpha }^2\right]=1$, which implies ${{\mu }_X=-\sigma }^2_X$ \cite{safari2008relay}. In order to thoroughly describe fading statistics of UWOC channels we should find out the dependency of log-amplitude variance to the ocean turbulence parameters.

 The scintillation index of a light wave is defined by \cite{andrews2001laser,korotkova2012light};
 \begin{equation}
{\sigma }^2_I\left(r,d_0,\lambda \right)=\frac{\left\langle I^2(r,d_0,\lambda )\right\rangle -{\left\langle I\left(r,d_0,\lambda \right)\right\rangle }^2}{{\left\langle I\left(r,d_0,\lambda \right)\right\rangle }^2},
\end{equation}
in which $I(r,d_0,\lambda )$ is the instantaneous intensity at a point with position vector $\left(r,d_0\right)=(x,y,d_0)$, where $d_0$ is the propagation distance and $\left \langle\cdot  \right \rangle$ denotes the long-time average.
Assuming weak turbulence, the scintillation index of plane and spherical waves can be obtained as \cite{andrews2001laser,korotkova2012light};
\begin{align} \label{SI}
&\sigma^2_I=8{\pi}^2k_0^2d_0\int_{0}^{1}\int_{0}^{\infty}\kappa{\Phi}_n(\kappa)\nonumber\\
&~~~~~~~~~\times\left \{1-{\rm cos}\left[\frac{d_0{\kappa}^2}{k_0}\xi\left(1-(1-\Theta)\xi\right)\right]  \right \}d\kappa d\xi,
\end{align}
in which $\Theta=1$ and $0$ for plane and spherical waves, respectively.
 $k_0={2\pi }/{\lambda }$ and $\kappa$ denote the wave number and scalar spatial frequency, respectively; {and ${\Phi }_n(\kappa )$ is the power spectrum of turbulent fluctuations given by Eq. (8) of \cite{korotkova2012light}. For lognormal turbulent channels the scintillation index relates to the log-amplitude variance as ${\sigma }^2_I=\exp(4{\sigma }^2_X)-1$ \cite{andrews2001laser}.
 In \cite{korotkova2012light,ata2014scintillations} the scintillation index of plane and spherical waves is numerically evaluated versus communication distance $d_0$ for various values of scintillation parameters and it is observed that, depending on the values of these parameters, the strong turbulence (which corresponds to $\sigma^2_I\geq 1$ \cite{andrews2001laser,korotkova2012light}) can occur at distances as long as $100$ {m} and as short as $10$ {m}, which impressively differs from atmospheric channels where strong turbulence distances are on the order of kilometers \cite{korotkova2012light}. Therefore, mitigating such a strong turbulence demands more attention.}

\subsection{MIMO UWOC System Model}
 We consider a UWOC system where the information signal is transmitted by $M$ transmitters, received by $N$ apertures, and combined using OC/EGC. As it is depicted in Fig. \ref{Fig1}, optical signal through propagation from the $i$th transmitter ${TX}_i$ to the $j$th receiver ${RX}_j$ experiences absorption, scattering, and turbulence, where in this paper absorption and scattering are modeled by fading-free impulse response of $h_{0,ij}(t)$ (as discussed in Section II-A) and turbulence is characterized by a multiplicative fading coefficient ${\alpha }^2_{ij}$, which has a lognormal distribution for weak oceanic turbulence (as explained in Section II-B). Therefore, the ${TX}_i$ to ${RX}_j$ channel has the aggregated impulse response of $h_{i,j}(t)={\alpha }^2_{ij}h_{0,ij}(t)$.
  
{We assume intensity-modulation direct-detection (IM/DD) with OOK signaling where the ``ON" state signal will be transmitted with the pulse shape $P(t)$.} In the special case of rectangular pulse, $P(t)$ can be represented as $P\left(t\right)=P\Pi \left(\frac{t-{T_b}/{2}}{T_b}\right)$, where $P$ is the transmitted power per bit ``$1$", $T_b$ is the bit duration time, and $\Pi(t)$ is a rectangular pulse with unit amplitude in the interval $[-1/2,1/2]$. Therefore, the transmitted signal can be expressed as;
\begin{equation}
S\left(t\right)=\sum^{\infty }_{k=-\infty }{b_kP\left(t-kT_b\right)},
\end{equation}
\noindent where $b_k\in \left\{0,1\right\}$ is the $k$th time slot transmitted bit. Hence, the received optical signal after propagating through the channel can be represented as;
\begin{equation} \label{7}
y\left(t\right)=S\left(t\right)*{\alpha }^2h_0(t)=\sum^{\infty}_{{k=-\infty}}{b_k{\alpha }^2\Gamma \left(t-kT_b\right)\ },
\end{equation}
in which $h_0(t)$ is the channel fading-free impulse response and $\Gamma \left(t\right)=h_0(t)*P(t)$, where $*$ denotes the convolution operator.

 When $M$ transmitters are used, in order to have a fair comparison with the single transmitter case, we assume that the total transmitted power for the ``ON" state signal is yet $P=\sum^M_{i=1}{P_i}$, where $P_i$ is the transmitted power by ${TX}_i$. In this regard, we denote the transmitted pulse from ${TX}_i$ by $P_{i}\left(t\right)=P_{i}\Pi \left(\frac{t-{T_b}/{2}}{T_b}\right)$ and also a sequence of data signal from ${TX}_i$ is denoted by $S_{i}\left(t\right)=\sum^{\infty }_{k=-\infty }{b_kP_{i}\left(t-kT_b\right)}$. This signal reaches ${RX}_j$ after propagation through the channel with impulse response ${\alpha }^2_{ij}h_{0,ij}(t)$. Hence, the transferred optical signal from ${TX}_i$ to ${RX}_j$ will be as;
\begin{equation} \label{8}
y_{i,j}\left(t\right)\!=\!S_{i}\left(t\right)*{\alpha }^2_{ij}h_{0,ij}(t)\!=\!\!\sum^{\infty }_{\substack{
   k=-\infty
  }}\!\!\!{b_k{\alpha }^2_{ij}{\Gamma }_{i,j}\left(t-kT_b\right)\ },
\end{equation}
where ${\Gamma }_{i,j}(t)=P_{i}\left(t\right)*h_{0,ij}\left(t\right)$. At the $j$th receiver, transmitted signals from all transmitters are captured, each with its own channel impulse response. In other words, received optical signal at the $j$th receiver is as $y_j\left(t\right)=\sum^M_{i=1}{y_{i,j}\left(t\right)}$ \cite{navidpour2007ber,tsiftsis2009optical}. Moreover, in order to have a fair comparison between multiple receivers and single receiver cases, we assume that the sum of all receiving apertures' areas in multiple receivers scheme is equal to the receiver aperture size of the single receiver scheme.

 We should emphasize that all of the expressions in the rest of this paper are based on ${\Gamma }_{i,j}(t)$, which is in terms of $P_{i}\left(t\right)$ and $h_{0,ij}\left(t\right)$. Therefore, our analytical expressions are applicable for any form of power allocation between different transmit apertures, any pulse shape of transmitted signal $P_{i}\left(t\right)$, and any channel model. However, our numerical results are based on equal power allocation for transmitters, i.e., $P_{i}={P}/{M},\ i=1,\dots ,M$, rectangular pulse for OOK signaling, MC-based simulated channel impulse response, and lognormal distribution for fading statistics. {Additionally, all of the derivations throughout the paper are for a general case from the transmitters and receivers structures point of view, i.e., all of the transmitters and receivers are located in arbitrary places. This generality is covered by considering a specific impulse response for any transmitter-to-receiver pair as $h_{i,j}(t)={\alpha }^2_{ij}h_{0,ij}(t)$. However, in our numerical results we assume that all of the transmitters and also all of the receivers are located with an equal separation distance in a line perpendicular to the transmission direction.}

\section{BER Analysis}
In this section, we analytically derive the exact and an upper bound BER expressions for both single-input single-output (SISO) and MIMO schemes, when either OC or EGC is used. Various noise components, i.e., background light, dark current, thermal noise, and signal-dependent shot noise all affect the system performance. Since these components are additive and independent of each other, in this section we model them as an equivalent noise component with Gaussian distribution \cite{lee2004part}. Also as it is shown in \cite{jamali2015ber}, the signal-dependent shot noise has a negligible effect with respect to the other noise components. Hence, it is amongst the other assumptions of this section to consider the noise variance independent to the incoming signal power. Moreover, we assume symbol-by-symbol processing at the receiver side, which is suboptimal in the presence of ISI \cite{einarsson1996principles}. In other words, the receiver integrates its output current over each $T_b$ seconds and then compares the result with an appropriate threshold to detect the received data bit. In this detection process, the availability of channel state information (CSI) is also assumed for threshold calculation.
\subsection{SISO UWOC Link}
{Based on Eq. \eqref{7}, the photodetector's $0$th time slot integrated current in SISO scheme can be expressed as};
\begin{align} \label{eq4}
r^{(b_0)}_{\rm SISO}=b_0{\alpha^2}\gamma^{(s)}+{\alpha^2}\sum_{k=-L}^{-1}b_k\gamma^{(I,k)}+v_{T_b},
\end{align}
where $\gamma^{(s)}=\boldmath{R}\int_{0}^{T_b}\Gamma(t)dt$, $\gamma^{(I,k)}=\boldmath{R}\int_{0}^{T_b}\Gamma(t-kT_b)dt=\boldmath{R}\int_{-kT_b}^{-(k-1)T_b}\Gamma(t)dt$, and $\boldmath{R}={\eta q}/{hf}$ is the photodetector's responsivity. {Moreover, $\eta$, $q$, $h$, $f$, and $L$ are the photodetector's quantum efficiency, electron's charge, Planck's constant, optical source frequency, and channel memory, respectively.} Physically, $\gamma^{(I,k\neq0)}$ refers to the ISI effect and $\gamma^{(I,k=0)}$ interprets the desired signal contribution, i.e., $\gamma^{(I,k=0)}=\gamma^{(s)}$. Furthermore, $v_{T_b}$ is the receiver integrated noise component which has a Gaussian distribution with mean zero and variance $\sigma^2_{T_b}$ \cite{lee2004part}. {{Note that based on the numerical results presented in \cite{tang2013temporal}, the channel correlation time is on the order of $10^{-5}$ to $10^{-2}$ seconds which implies that thousands up to millions of consecutive bits have the same fading coefficient. Therefore, we have adopted the same fading coefficient for all of the consecutive bits in Eq. \eqref{eq4}.}}

Assuming the availability of CSI, the receiver compares its integrated current over each $T_b$ seconds with an appropriate threshold, i.e., with $\tilde{T}={\alpha^2}\gamma^{(s)}/2$. Therefore, the conditional probabilities of error when bits ``$1$" and ``$0$" are transmitted can respectively be obtained as;
\begin{align} \label{eq5}
P^{\rm SISO}_{be|1,{\alpha},b_k}&=\Pr(r^{(b_0)}_{\rm SISO}\leq \tilde{T}|b_0=1)\nonumber\\
&=Q\left(\frac{{\alpha^2}\left[\gamma^{(s)}/2+\sum_{k=-L}^{-1}b_k\gamma^{(I,k)}\right]}{\sigma_{T_b}}\right),
\end{align}
\begin{align} \label{eq6}
P^{\rm SISO}_{be|0,{\alpha},b_k}&=\Pr(r^{(b_0)}_{\rm SISO}\geq \tilde{T}|b_0=0)\nonumber\\
&=Q\left(\frac{{\alpha^2}\left[\gamma^{(s)}/2-\sum_{k=-L}^{-1}b_k\gamma^{(I,k)}\right]}{\sigma_{T_b}}\right),
\end{align}
where $Q\left(x\right)=({1}/{\sqrt{2\pi }})\int^{\infty }_x{{\rm exp}({-{y^2}/{2}})}dy$ is the Gaussian-Q function. Then the final BER can be obtained by averaging the conditional BER $P^{\rm SISO}_{be|{\alpha},b_k}=\frac{1}{2}P^{\rm SISO}_{be|0,{\alpha},b_k}+\frac{1}{2}P^{\rm SISO}_{be|1,{\alpha},b_k}$ over fading coefficient ${\alpha}$ and all $2^L$ possible data sequences for $b_k$s as;
\begin{align}\label{eq7}
P^{\rm SISO}_{be}=\frac{1}{2^L}\sum_{b_k}\int_{0}^{\infty}P^{\rm SISO}_{be|{\alpha},b_k}f_{{\alpha}}({\alpha})d{\alpha}.
\end{align}

The forms of Eqs. \eqref{eq5} and \eqref{eq6} suggest an upper bound on the system BER, from the ISI point of view. In other words, $b_{k\neq0}=0$ maximizes \eqref{eq5}, while \eqref{eq6} has its maximum value for $b_{k\neq0}=1$. Indeed, when data bit ``$0$" is sent{,} the worst effect of ISI occurs when all of the surrounding bits are ``$1$" (i.e., when $b_{k\neq0}=1$), and vice versa \cite{einarsson1996principles}. Regarding these special sequences, the upper bound on the BER of SISO UWOC system can be evaluated as;
\begin{align}\label{eq8}
& P^{\rm SISO}_{be,{\rm UB}}=\frac{1}{2}\int_{0}^{\infty}\Bigg[ Q\left(\frac{{\alpha^2}\gamma^{(s)}}{2\sigma_{T_b}}\right)+\nonumber\\
&~~~~~~~~ Q\left(\frac{{\alpha^2}\left[\gamma^{(s)}/2-\sum_{k=-L}^{-1}\gamma^{(I,k)}\right]}{\sigma_{T_b}}\right)\Bigg]f_{\alpha}(\alpha)d\alpha.
\end{align}

The averaging over fading coefficient in \eqref{eq7} and \eqref{eq8} involves integrals of the form $\int_{0}^{\infty}Q(C{\alpha^2})f_{\alpha}(\alpha)d\alpha$, where $C$ is a constant, e.g., $C=\gamma^{(s)}/2\sigma_{T_b}$ in the first integral of \eqref{eq8}. Such integrals can effectively be calculated using Gauss-Hermite quadrature formula [37, Eq. (25.4.46)] as;
\begin{align} \label{eq9}
&\int_{0}^{\infty}Q(C{\alpha^2})f_{\alpha}(\alpha)d\alpha&\nonumber\\
&~~~~~=\int_{-\infty}^{\infty}Q(Ce^{2x})\frac{1}{\sqrt{2\pi\sigma^2_X}}\exp\left(-\frac{(x-\mu_X)^2}{2\sigma^2_X}\right)dx\nonumber\\
&~~~~~\approx \frac{1}{\sqrt{\pi}}\sum_{q=1}^{U}w_qQ\left(C \exp\left(2x_q\sqrt{2\sigma^2_X}+2\mu_X\right)\right),
\end{align}
in which $U$ is the order of approximation, $w_q,~q=1,2,...,U$, are weights of the $U$th-order approximation and $x_q$ is the $q$th zero of the $U$th-order Hermite polynomial, $H_U(x)$ \cite{navidpour2007ber,abramowitz1970handbook}.
\subsection{MIMO UWOC Link with OC}
Relying on \eqref{8}, the integrated current of the $j$th receiver can be expressed as;
\begin{align} \label{Eq.III-B1}
r^{(b_0)}_{j}=b_0\sum_{i=1}^{M}\!{\alpha}^2_{ij}\gamma^{(s)}_{i,j}\!+\!\sum_{i=1}^{M}\!{\alpha}^2_{ij}\!\!\!\sum_{k=-L_{ij}}^{-1}\!\!\!\!b_k\gamma^{(I,k)}_{i,j}+v^{(j)}_{T_b},
\end{align}
where $\gamma^{(s)}_{i,j}=\boldmath{R}\int_{0}^{T_b}\Gamma_{i,j}(t)dt$, $\gamma^{(I,k)}_{i,j}=\boldmath{R}\int_{0}^{T_b}\Gamma_{i,j}(t-kT_b)dt=\boldmath{R}\int_{-kT_b}^{-(k-1)T_b}\Gamma_{i,j}(t)dt$, $L_{i,j}$ is the memory of the channel between the $i$th transmitter and $j$th receiver, and $v^{(j)}_{T_b}$ is the $j$th receiver integrated noise which has a Gaussian distribution with mean zero and variance $\sigma^2_{T_b}$. {It is {worth mentioning} that while EGC simply adds the output of each receiving branch (with equal gain) to construct the combined output, the OC applies maximum a posteriori (MAP) detection rule (which in the case of symbols with equal probability simplifies to ML rule) to obtain the optimal combining policy.}

Using the derived decision rule in Appendix A, we find the ``ON" and ``OFF" states conditional error probabilities as Eqs. \eqref{Eq.III-B6} and \eqref{Eq.III-B7}, respectively, shown at the top of the next page.
\begin{figure*}[!t]
\normalsize
\setcounter{equation}{13}
\begin{align}\label{Eq.III-B6}
P^{\rm MIMO}_{be,{\rm OC}|1,\vec{\boldsymbol{\alpha}},b_k}&=\Pr\left(\sum_{j=1}^{N}2r_j\sum_{i=1}^{M}\alpha_{ij}^2\gamma^{(s)}_{i,j}\leq\sum_{j=1}^{N}\left(\sum_{i=1}^{M}\alpha_{ij}^2\gamma^{(s)}_{i,j}\right)^{2}{\Big |}r_j=\sum_{i=1}^{M}{\alpha}^2_{ij}\gamma^{(s)}_{i,j}+\sum_{i=1}^{M}{\alpha}^2_{ij}\sum_{k=-L_{ij}}^{-1}b_k\gamma^{(I,k)}_{i,j}+v^{(j)}_{T_b},\vec{\boldsymbol{\alpha}},b_k\right)\nonumber\\
&=\Pr\left(\sum_{j=1}^{N}\left[2v^{(j)}_{T_b}\sum_{i=1}^{M}\alpha_{ij}^2\gamma^{(s)}_{i,j}\right]\leq -\sum_{j=1}^{N}\sum_{i'=1}^{M}{\alpha}^2_{i'j}\gamma^{(s)}_{i',j}\sum_{i=1}^{M}{\alpha}^2_{ij}\left(\gamma^{(s)}_{i,j}+2\sum_{k=-L_{ij}}^{-1}b_k\gamma^{(I,k)}_{i,j}\right)\right)
\nonumber\\
&=Q\left(\frac{\sum_{j=1}^{N}\sum_{i'=1}^{M}{\alpha}^2_{i'j}\gamma^{(s)}_{i',j}\sum_{i=1}^{M}{\alpha}^2_{ij}\left(\gamma^{(s)}_{i,j}+2\sum_{k=-L_{ij}}^{-1}b_k\gamma^{(I,k)}_{i,j}\right)}{2\sigma_{T_b}\sqrt{\sum_{j=1}^{N}\left(\sum_{i=1}^{M}\alpha_{ij}^2\gamma^{(s)}_{i,j}\right)^2}}\right).
\end{align}
\hrulefill
\begin{align} \label{Eq.III-B7}
P^{\rm MIMO}_{be,{\rm OC}|0,\vec{\boldsymbol{\alpha}},b_k}=Q\left(\frac{\sum_{j=1}^{N}\sum_{i'=1}^{M}{\alpha}^2_{i'j}\gamma^{(s)}_{i'\!,j}\sum_{i=1}^{M}{\alpha}^2_{ij}\left(\gamma^{(s)}_{i,j}-2\sum_{k=-L_{ij}}^{-1}b_k\gamma^{(I,k)}_{i,j}\right)}{2\sigma_{T_b}\sqrt{\sum_{j=1}^{N}\left(\sum_{i=1}^{M}\alpha_{ij}^2\gamma^{(s)}_{i,j}\right)^2}}\right).
\end{align}
\hrulefill
\end{figure*}
Assuming the maximum channel memory as $L_{\rm max}=\max\{L_{11},L_{12},...,L_{MN}\}$, the average BER of MIMO UWOC system can be obtained by averaging over fading coefficients vector $\vec{\boldsymbol{\alpha}}$ (through an ($M\times N$)-dimensional integral) as well as averaging over all $2^{L_{\rm max}}$ possible sequences for $b_k$s, i.e.,
\begin{align}\label{Eq.III-B8}
\!P^{\rm MIMO}_{be,{\rm OC}}=\frac{1}{2^{L_{\rm max}}}\sum_{b_k}\!\int_{\vec{\boldsymbol{\alpha}}}\frac{1}{2}\big[& P^{\rm MIMO}_{be,{\rm OC}|1,{\vec{\boldsymbol{\alpha}}},b_k}+\nonumber\\&P^{\rm MIMO}_{be,{\rm OC}|0,{\vec{\boldsymbol{\alpha}}},b_k}\big]f_{\vec{\boldsymbol{\alpha}}}(\vec{\boldsymbol{\alpha}})d\vec{\boldsymbol{\alpha}},\!
\end{align}
where $f_{\vec{\boldsymbol{\alpha}}}(\vec{\boldsymbol{\alpha}})$ is the joint PDF of fading coefficients in $\vec{\boldsymbol{\alpha}}$.

\begin{figure*}[!t]
\normalsize
\setcounter{equation}{16}
\begin{align}\label{Eq.III-B9}
P^{\rm MIMO}_{be,{\rm OC,UB}}\!\!=\!\int_{\vec{\boldsymbol{\alpha}}}\!\frac{1}{2}\!\left[Q\!\left(\frac{\sqrt{\sum_{j=1}^{N}\!\left(\sum_{i=1}^{M}\!\alpha_{ij}^2\gamma^{(s)}_{i,j}\!\right)^2}}{2\sigma_{T_b}}\right)
\!+\! Q\!\left(\!\frac{\sum_{j=1}^{N}\!\sum_{i'=1}^{M}\!{\alpha}^2_{i'j}\gamma^{(s)}_{i'\!,j}\!\sum_{i=1}^{M}\!{\alpha}^2_{ij}\!\left(\!\gamma^{(s)}_{i,j}\!-\!2\sum_{k=-L_{ij}}^{-1}\gamma^{(I,k)}_{i,j}\right)}{2\sigma_{T_b}\sqrt{\sum_{j=1}^{N}\left(\sum_{i=1}^{M}\alpha_{ij}^2\gamma^{(s)}_{i,j}\right)^2}}\!\right)\!\right]\!f_{\vec{\boldsymbol{\alpha}}}(\vec{\boldsymbol{\alpha}})d\vec{\boldsymbol{\alpha}}.
\end{align}
\hrulefill
\end{figure*}
Furthermore, considering the transmitted data sequences as $b_{k\neq0}=1$ for $b_0=0$ and $b_{k\neq0}=0$ for $b_0=1$, the upper bound on the BER of MIMO UWOC system can be obtained as Eq. \eqref{Eq.III-B9}, shown at the top of the next page.
Moreover, as it is shown in Appendix B, ($M\times N$)-dimensional integrals in \eqref{Eq.III-B8} and \eqref{Eq.III-B9} can effectively be calculated by ($M\times N$)-dimensional series using Gauss-Hermite quadrature formula.

It is worth mentioning that for transmitter diversity ($N=1$) the conditional BER expressions in \eqref{Eq.III-B6} and \eqref{Eq.III-B7} simplify to;
\begin{align}\label{Eq.III-B10}
& P^{\rm MISO}_{be|b_0,\vec{\boldsymbol{\alpha}},b_k}=\!\nonumber\\&\!Q\left({{\underset{i=1}{\overset{M}{\sum}}}{\alpha^2_{i1}}\!\!\left[\gamma^{(s)}_{i,1}\!+\!(-1)^{b_0+1}\!\!\!\!\!{\underset{~k=-L_{i1}}{\overset{-1}{\sum}}}\!\!\!\!2b_k\gamma^{(I,k)}_{i,1}\!\right]}{\Big/}{2\sigma_{T_b}}\!\right),
\end{align}
which can be reformulated as $P^{\rm MISO}_{be|b_0,\vec{\boldsymbol{\alpha}},b_k}=Q\left(\sum_{i=1}^{M}\alpha^2_{i1}G^{(b_0)}_{i,1}\right)$, where $G^{(b_0)}_{i,1}=[\gamma^{(s)}_{i,1}+(-1)^{b_0+1}\sum_{k=-L_{i1}}^{-1}2b_k\gamma^{(I,k)}_{i,1}]/2\sigma_{T_b}$. The weighted sum of RVs in \eqref{Eq.III-B10} can be approximated by an equivalent RV, using moment matching method \cite{fenton1960sum}. Therefore, we can approximate the conditional BER of \eqref{Eq.III-B10} as $P^{\rm MISO}_{be|b_0,\vec{\boldsymbol{\alpha}},b_k}\approx Q\left(G^{(b_0)}_{M}\right)$, where $G^{(b_0)}_{M}$ is the equivalent RV resulted from the approximation to the weighted sum of RVs, i.e., $G^{(b_0)}_{M}\approx \sum_{i=1}^{M}\alpha^2_{i1}G^{(b_0)}_{i,1}$. In the special case of lognormal fading, the equivalent lognormal RV, $G^{(b_0)}_{M}=\exp(2z^{(b_0)})$, has the log-amplitude mean and variance of;
\begin{align}
      &\mu_{z^{(b_0)}}=\frac{1}{2}{\rm ln}\bigg(\sum_{i=1}^MG^{(b_0)}_{i,1}\bigg)-\sigma^2_{z^{(b_0)}},\label{Eq.III-B11}\\
       & \sigma^2_{z^{(b_0)}}=\frac{1}{4}{\rm ln}\left(1+\frac{\sum_{i=1}^M\left(G^{(b_0)}_{i,1}\right)^2\left(e^{4\sigma^2_{X_{i1}}}-1\right)}{\left(\sum_{i=1}^MG^{(b_0)}_{i,1}\right)^2}\right),\label{Eq.III-B12}
       \end{align}
respectively \cite{safari2008relay}. Hence, in the case of transmitter diversity the average BER can approximately be evaluated with a one-dimensional integral which can also be calculated using Gauss-Hermite quadrature formula as a one-dimensional series.
\subsection{MIMO UWOC Link with EGC}
When EGC is used, the integrated current of the receiver output, based on Eq. \eqref{8}, can be expressed as;
\begin{align} \label{Eq.III-C1}
r^{(b_0)}_{\rm MIMO}\!=\!b_0\!\sum_{j=1}^{N}\sum_{i=1}^{M}\!{\alpha}^2_{ij}\gamma^{(s)}_{i,j}\!+\!\!\sum_{j=1}^{N}\sum_{i=1}^{M}\!{\alpha}^2_{ij}\!\!\!\!\!\sum_{k=-L_{ij}}^{-1}\!\!b_k\gamma^{(I,k)}_{i,j}\!+v^{(N)}_{T_b},
\end{align}
where $v^{(N)}_{T_b}$ is the integrated combined noise component which has a Gaussian distribution with mean zero and variance $N\sigma^2_{T_b}$ \cite{jamali2015ber}. 

Based on \eqref{Eq.III-C1} and the availability of CSI, the receiver selects the threshold value as $\tilde{T}=\sum_{j=1}^{N}\sum_{i=1}^{M}\alpha^2_{ij}\gamma^{(s)}_{i,j}/2$. Pursuing similar steps as Section III-B results to \eqref{Eq.III-C2} for the conditional BER.
As expected, \eqref{Eq.III-C2} simplifies to \eqref{Eq.III-B10} for {MISO} scheme.
Finally, the average BER can be evaluated similar to \eqref{Eq.III-B8}. Also the upper bound on the BER of MIMO UWOC system with EGC can be expressed as Eq. \eqref{Eq.III-C3}, shown at the top of the next page.

\begin{figure*}[!t]
\normalsize
\setcounter{equation}{21}
\begin{align} \label{Eq.III-C2}
P^{\rm MIMO}_{be,{\rm EGC}|b_0,\vec{\boldsymbol{\alpha}},b_k}=Q\left(\frac{\sum_{j=1}^{N}\sum_{i=1}^{M}\alpha^2_{ij}\gamma^{(s)}_{i,j}+(-1)^{b_0+1}\sum_{j=1}^{N}\sum_{i=1}^{M}\alpha^2_{ij}\sum_{k=-L_{ij}}^{-1}2b_k\gamma^{(I,k)}_{i,j}}{2\sqrt{N}\sigma_{T_b}}\right).
\end{align}
\hrulefill
\begin{align}\label{Eq.III-C3}
P^{\rm MIMO}_{be,{\rm EGC,UB}}=\int_{\vec{\boldsymbol{\alpha}}}\frac{1}{2}\Bigg[
Q\left(\frac{\sum_{j=1}^{N}\sum_{i=1}^{M}\alpha^2_{ij}\gamma^{(s)}_{i,j}}{2\sqrt{N}\sigma_{T_b}}\right)
+
Q\left(\frac{\sum_{j=1}^{N}\sum_{i=1}^{M}\alpha^2_{ij}\left[\gamma^{(s)}_{i,j}-2\sum_{k=-L_{ij}}^{-1}\gamma^{(I,k)}_{i,j}\right]}{2\sqrt{N}\sigma_{T_b}}\right)
\Bigg]f_{\vec{\boldsymbol{\alpha}}}(\vec{\boldsymbol{\alpha}})d\vec{\boldsymbol{\alpha}}.
\end{align}
\hrulefill
\end{figure*}

It is worth noting that the numerator of \eqref{Eq.III-C2} can be approximated as $\zeta^{(b_0)}\approx\sum_{j=1}^{N}\sum_{i=1}^{M}D^{(b_0)}_{i,j}\alpha^2_{ij}$, where the weight coefficients are defined as $D^{(b_0)}_{i,j}=\gamma^{(s)}_{i,j}+(-1)^{b_0+1}\sum_{k=-L_{ij}}^{-1}2b_k\gamma^{(I,k)}_{i,j}$. Similar to \eqref{Eq.III-B11} and \eqref{Eq.III-B12}, statistics of the equivalent lognormal RV $\zeta^{(b_0)}$, which is resulted from weighted sum of $M\times N$ RVs, can be obtained and then averaging over fading coefficients reduces to the one-dimensional integral of;
\begin{align} \label{Eq.III-C4}
P^{\rm MIMO}_{be,{\rm EGC}|b_0,b_k}\approx\int_{0}^{\infty}Q\left(\frac{\zeta^{(b_0)}}{2\sqrt{N}\sigma_{T_b}}\right) f_{\zeta^{(b_0)}}(\zeta^{(b_0)})d\zeta^{(b_0)},
\end{align} 
which can also effectively be calculated using Eq. \eqref{eq9}.

 \section{BER Evaluation Using Photon-Counting Methods}
  In this section, we derive the required expressions for the system BER using photon-counting approach. Moreover, in this section signal-dependent shot noise, dark current, and background light all are considered with Poisson distribution, while thermal noise is assumed to be Gaussian distributed \cite{einarsson1996principles}.
  To evaluate the BER, we can apply either saddle-point approximation or Gaussian approximation which is simpler but negligibly less accurate than saddle-point approximation.
    Based on saddle-point approximation the system BER can be obtained as $P_{be}=\frac{1}{2}\left[q_+\left(\beta \right)+q_-(\beta )\right]$,
in which $q_+(\beta )$ and $q_-(\beta )$ are probabilities of error when bits ``$0$'' and ``$1$'' are sent, respectively, i.e.,
 \begin{align} \label{10}
 & q_+\left(\beta \right)=\Pr\left(u>\beta|zero\right)\approx \frac{{\rm exp}\left[{\Phi }_0(s_0)\right]}{\sqrt{2\pi {\Phi }^{''}_0(s_0)}},\nonumber\\
 & q_-\left(\beta \right)=\Pr\left(u\leq\beta|one\right)\approx \frac{{\rm exp}\left[{\Phi }_1(s_1)\right]}{\sqrt{2\pi {\Phi }^{''}_1(s_1)}},\nonumber\\
 & {\Phi }_{b_0}\left(s\right)={\rm ln}\left[{\Psi }_{u^{(b_0)}}\left(s\right)\right]-s\beta -{\rm ln}\left|s\right|,~~~b_0=0,1,\
 \end{align}
 \noindent where $u$ is the photoelectrons count at the receiver output and ${\Psi }_{u^{(b_0)}}\left(s\right)$ is the receiver output moment generating function (MGF) when bit ``$b_0$'' is sent. Also $s_0$ is the positive and real root of ${\Phi }^{'}_0(s)$, i.e., ${\Phi }^{'}_0(s_0)=0$ and $s_1$ is the negative and real root of ${\Phi }^{'}_1(s)$, i.e., ${\Phi }^{'}_1(s_1)=0$; and $\beta $ is the receiver optimum threshold and will be chosen such that it minimizes the error probability, i.e., ${dP_{be}}/{d\beta }=0$.
 As an another approach to evaluate the system BER, Gaussian approximation is very fast and computationally efficient, yet not as accurate as saddle-point approximation, but yields an acceptable estimate of the system error rate particularly for BER values smaller than $0.1$ \cite{einarsson1996principles}. Indeed, when the receiver output is as $u=\boldsymbol{N}+\xi $, where $\boldsymbol{N}$ is a Poisson distributed RV with mean $m^{(b_0)}$ for {the} transmitted bit ``$b_0$" and $\xi$ is a Gaussian distributed RV with mean zero and variance ${\sigma }^2$, Gaussian approximation which approximates $\boldsymbol{N}$ as a Gaussian distributed RV with equal mean and variance results to the following equation for the system BER \cite{einarsson1996principles};
\begin{align}\label{11}
  P_{be}=Q\left(\frac{m^{(1)}-m^{(0)}}{\sqrt{m^{(1)}+{\sigma }^2}+\sqrt{m^{(0)}+{\sigma }^2}}\right).
 \end{align} 
  In this section, the required expressions for both the exact and the upper bound BER evaluations using either saddle-point or Gaussian approximation are presented, when EGC is used at the receiver side.
 \subsection{SISO Configuration}
Based on Eq. \eqref{7}, the photo-detected signal generated by the integrate-and-dump circuit of the SISO receiver can be expressed as;
 \begin{equation} \label{r^{b_0}}
 u^{(b_0)}_{\rm SISO}=y^{(b_0)}_{\rm SISO}+v_{th},
 \end{equation}
 where $v_{th}$ corresponds to the receiver integrated thermal noise and is a Gaussian distributed RV with mean zero and variance $\sigma^2_{th}={2K_bT_rT_b}/{(R_Lq^2)}$, where $K_b$, $T_r$, and $R_L$ are Boltzmann's constant, the receiver equivalent temperature, and load resistance, respectively   \cite{jazayerifar2006atmospheric}. Conditioned on $\left \{ b_k \right \}_{k=-L}^{-1}$ and $\alpha$, $y^{(b_0)}_{\rm SISO}$ is a Poisson distributed RV with mean $m^{(b_0)}_{\rm SISO}$ as;
 \begin{align} \label{m^{b_0}}
 & m^{(b_0)}_{\rm SISO}=\frac{\eta {\alpha }^2}{hf}\sum^{0}_{{k=-L}}\!\!{b_k\!\int^{T_b}_0{\!\!\!\Gamma \left(t-kT_b\right)dt}}\!+\!(n_b\!+\!n_d)T_b,
  \end{align}
 in which ${n}_b$ and ${n}_d$ are mean count rates of Poisson distributed background radiation and dark current noise, respectively. 
%

  As it is shown in Appendix C, conditioned on $\alpha$ the receiver output MGF can be expressed as;
  \begin{align} \label{SISO_MGF}
  {\Psi }_{{u^{(b_0)}_{\rm SISO}}|\alpha }(s)=&\exp\left({\frac{s^2{\sigma }^2_{th}}{2}}\!+\!{\left[m_{\rm SISO}^{(bd)}+b_0{\alpha}^2m^{(s)}\right]\left(e^s-1\right)}\right)\nonumber\\
  &
  \!\!\times\!\!\prod_{k=-L}^{-1}\left[\frac{1+{\rm exp}\left({\alpha}^2 m^{(I,k)}\left(e^s\!-\!1\right)\right)}{2}\right],
  \end{align}
  in which $m_{\rm SISO}^{(bd)}=\left(n_b+n_d\right)T_b$, $m^{(s)}=\frac{\eta}{hf}\int^{T_b}_0{\Gamma(t)dt}$, and $m^{(I,k)}=\frac{\eta}{hf}\int^{T_b}_0{\Gamma(t-kT_b)dt}=\frac{\eta}{hf}\int^{(-k+1)T_b}_{-kT_b}{\Gamma(t)dt}$.
 Moreover, assuming the transmitted data sequences as $b_{k\neq0}=1$ for $b_0=0$ and $b_{k\neq0}=0$ for $b_0=1$, MGF of the receiver output for evaluation of upper bound on the BER of SISO UWOC system can be obtained as;
 \begin{align} \label{MGF_SISO_UB}
{\Psi }^{\rm UB}_{u^{(b_0)}_{\rm SISO}|\alpha}\left(s\right)=\exp\Bigg(&\frac{s^2{\sigma }^2_{th}}{2}+\Bigg[m_{\rm SISO}^{(bd)}+b_0\alpha^2m^{(s)}\nonumber\\
&+\sum_{k=-L}^{-1}\overline{b_0}\alpha^2m^{(I,k)}\Bigg](e^s-1)\Bigg),
 \end{align}
 where $\overline{b_0}=1-b_0$. Inserting \eqref{SISO_MGF} and \eqref{MGF_SISO_UB} in \eqref{10} results into {the} conditional BER, $P_{be|\alpha}$, and the final BER can then be obtained by averaging over {the} fading coefficient $\alpha$.
 \subsection{MIMO Configuration with EGC}
  In this scheme, each of $N$ receiving apertures receives the sum of all transmitters signals.
%
 At the receiver side, each of these $N$ received signals passes through its receiver photodetector and different types of noises are added to each output. Therefore, the photo-detected signal at the $j$th receiver generated by integrate-and-dump circuit can be expressed as $u^{(b_0)}_{j}=y^{(b_0)}_{j}+{v_{th,j}}$, where ${v_{th,j}}$ is a Gaussian distributed RV with mean zero and variance ${{\sigma}^2_{th,j}}={{\sigma}^2_{th}}$ corresponding to the integrated thermal noise of the $j$th receiver and $y^{(b_0)}_{j}$ conditioned on $\left \{ b_k \right \}_{k=-L_{ij}}^{-1}$ and $\left \{\alpha_{ij}\right \}_{i=1}^{M}$ is a Poisson distributed RV with mean;
\begin{align} \label{m^{b_0}_j}
  m^{(b_0)}_{j}\!=\!\frac{\eta }{hf}\!\sum^M_{i=1}\!\!\!\!\!\!\sum^{0}_{~~~k=-L_{ij}}{{\!\!\!\!\!\!\!{\alpha }^2_{ij}b_k\!\!\int^{T_b}_0{{\!\!\!\Gamma }_{i,j}\left(t\!-\!kT_b\right)\!dt\!}}}+\!({n}_{d,j}\!+\!{n}_{b,j})T_b,
 \end{align} 
 where $n_{b,j}$ and $n_{d,j}$ are the mean count rates of Poisson distributed background radiation and dark current noise of the $j$th receiver, respectively. As it is demonstrated in Appendix D, MGF of the receiver output in MIMO scheme conditioned on fading coefficients vector $\vec{\boldsymbol{\alpha}}=(\alpha_{11},\alpha_{12},...,\alpha_{MN})$ can be expressed as Eq. \eqref{MGF_MIMO},\footnote{Note that each of the receivers introduces a Gaussian distributed thermal noise with mean zero and variance ${{\sigma}^2_{th,j}}={{\sigma}^2_{th}}$ and a Poisson distributed dark current with mean count of $n_{d,j}T_b=n_dT_b$. But mean of the Poisson distributed background noise is proportional to the receiver aperture size and we assumed that the sum of all receiving apertures is identical to the aperture size of MISO scheme, which implies that $\sum_{j=1}^{N}n_{b,j}=n_b$.}
\begin{figure*}[!t]
\normalsize
\setcounter{equation}{31}
\begin{align} \label{MGF_MIMO}
{\Psi }^{{\rm{EGC}}}_{{u^{(b_0)}_{\rm MIMO}}{|}\vec{\boldsymbol{\alpha}}}\left(s\right)\!=\!{\rm exp}\left(\!{\frac{{N\sigma }^2_{th}}{2}s^2}\!\!+\!\!\left[{m_{\rm MIMO}^{(bd)}}\!+\!\sum_{j=1}^{N}\!\sum_{i=1}^{M}b_0{\alpha^2_{ij}}{m_{i,j}^{(s)}}\right]\!\left(e^s\!-\!1\right)\right)\!\times\!\prod_{j=1}^{N}\!\!\!\prod_{~~k=-{L_{\rm max}}}^{-1}\!\!\!\!\!\frac{1}{2}\!\left[1\!+\!\prod_{i=1}^{M}\!{\rm exp}\left({\alpha^2_{ij}}m_{i,j}^{(I,k)}\!\left(e^s\!-\!1\right)\right)\right].
 \end{align}
\hrulefill
\begin{align} \label{MGF_MIMO_UB}
  {\Psi }^{{\rm{EGC},\rm{UB}}}_{{u^{(b_0)}_{\rm MIMO}}|{\vec{\boldsymbol{\alpha}}} }\left(s\right)={\rm exp}\Bigg({\frac{{N\sigma }^2_{th}}{2}s^2}+\left[{m_{\rm MIMO}^{(bd)}}+\sum_{j=1}^{N}\sum_{i=1}^{M}{\alpha^2_{ij}}\left(b_0{m_{i,j}^{(s)}}+\overline{b_0}\sum_{k=-L_{ij}}^{-1}m_{i,j}^{(I,k)}\right)\right]\left(e^s-1\right)\Bigg).
  \end{align}
  \hrulefill
\end{figure*}
in which $m_{\rm MIMO}^{(bd)}=\left(n_b+Nn_d\right)T_b$, $m^{(s)}_{i,j}=\frac{\eta}{hf}\int^{T_b}_0{\Gamma_{i,j}(t)dt}$, and $m_{i,j}^{(I,k)}=\frac{\eta}{hf}\int^{T_b}_0{\Gamma_{i,j}(t-kT_b)dt}=\frac{\eta}{hf}\int^{(-k+1)T_b}_{-kT_b}{\Gamma_{i,j}(t)dt}$. Furthermore, MGF of the receiver output for {the} evaluation of the upper bound on the BER of MIMO UWOC system can be expressed as Eq. \eqref{MGF_MIMO_UB}, shown at the top of the next page.
 
  We should emphasize that extracting the output MGFs for MISO and single-input multiple-output (SIMO) schemes is straightforward by respectively substituting $N=1$ and $M=1$ in \eqref{MGF_MIMO} and \eqref{MGF_MIMO_UB}. {Note that for these cases $m_{\rm MISO}^{(bd)}=m_{\rm SISO}^{(bd)}$ and $m_{\rm SIMO}^{(bd)}=m_{\rm MIMO}^{(bd)}$.}
Eventually, using saddle-point approximation the conditional BER $P_{be|\vec{\boldsymbol{\alpha}}}$ can be achieved by inserting \eqref{MGF_MIMO} and \eqref{MGF_MIMO_UB} in \eqref{10}. The final BER can then be evaluated by averaging over $\vec{\boldsymbol{\alpha}}$ as $P_{be}=\int_{\vec{\boldsymbol{\alpha}}}{P_{be|\vec{\boldsymbol{\alpha}}}}{f_{\vec{\boldsymbol{\alpha}}}\left(\vec{\boldsymbol{\alpha}}\right)}d\vec{\boldsymbol{\alpha}}$.
 
 With respect to the above complex expressions, using saddle-point approximation for BER evaluation may be difficult and computationally time-consuming, since it needs to solve some complicated equations for which their complexity increases as ISI (or equivalently $L_{\rm max}$ in \eqref{MGF_MIMO}) increases. But \eqref{11} suggests that using Gaussian approximation is simple and computationally fast. 
It can easily be shown that conditioned on $b_k$ the receiver output signal is the sum of a Gaussian and a Poisson RVs; therefore, Gaussian approximation can be applied to evaluate the average BER conditioned on $\vec{\boldsymbol{\alpha}}$ and $b_k$, i.e., $P_{be|\vec{\boldsymbol{\alpha}},b_k}$.
The Gaussian distributed RV has mean zero and variance ${N\sigma }^2_{th}$. And the Poisson distributed RV has mean of;
\begin{align}\label{M-gaus:4}
 m^{(b_0)}_{\rm MIMO}\!=\!m_{\rm MIMO}^{(bd)}\!+\!\!\sum_{j=1}^{N}\sum_{i=1}^{M}\left[b_0{\alpha^2_{ij}}{m_{i,j}^{(s)}}\!+\!\!\!\!\sum_{k=-L_{ij}}^{-1}\!\!\!\!b_k{\alpha^2_{ij}}{m_{i,j}^{(I,k)}}\right].
\end{align}
  Moreover, mean of the Poisson distributed RV for {the} evaluation of {the} upper bound on the BER of UWOC system can easily be obtained by assuming the transmitted data sequences as $b_{k\neq0}=1$ for $b_0=0$ and $b_{k\neq0}=0$ for $b_0=1$.

Using \eqref{11} and \eqref{M-gaus:4} the conditional BER, $P_{be|\vec{\boldsymbol{\alpha}},b_k}$, can easily be evaluated based on Gaussian approximation. Moreover, to obtain $P_{be|\vec{\boldsymbol{\alpha}},b_k}$ based on saddle-point approximation, the simplified form of saddle-point approximation [36, Eqs. (5.73)-(5.79)] can be applied to \eqref{M-gaus:4}. Subsequently, $P_{be|b_k}$ can be obtained through an ($M\times N$)-dimensional integration as $P_{be|b_k}=\int_{\vec{\boldsymbol{\alpha}}}{P_{be|\vec{\boldsymbol{\alpha}},b_k}}{f_{\vec{\boldsymbol{\alpha}}}\left(\vec{\boldsymbol{\alpha}}\right)}d\vec{\boldsymbol{\alpha}}$ which yet demands excessive computational time, especially for large number of links. Nevertheless, we can reformulate \eqref{M-gaus:4} as $m^{(b_0)}_{\rm MIMO}=m_{\rm MIMO}^{(bd)}+\sum_{j=1}^{N}\sum_{i=1}^{M}\left[\tau^{(b_0)}_{i,j}{\alpha^2_{ij}}\right]$, where $\tau^{(b_0)}_{i,j}=b_0m_{i,j}^{(s)}+\sum_{k=-L_{ij}}^{-1}b_km_{i,j}^{(I,k)}$.
Hence, we can approximate \eqref{M-gaus:4} as ${m}^{(b_0)}_{\rm MIMO}\approx{\tilde{m}}^{(b_0)}_{\rm MIMO}=m_{\rm MIMO}^{(bd)}+\vartheta^{(b_0)}$, where ${\tilde{m}}^{(b_0)}_{\rm MIMO}$ is the approximated version of \eqref{M-gaus:4} and $\vartheta^{(b_0)}\approx\sum_{j=1}^{N}\sum_{i=1}^{M}\tau^{(b_0)}_{i,j}{\alpha^2_{ij}}$, i.e., the weighted sum of $M\times N$ RVs.
   In the special case of weak oceanic turbulence the equivalent lognormal RV, $\vartheta^{(b_0)}={\rm exp}(2z^{(b_0)})$, has the following log-amplitude mean and variance, respectively \cite{safari2008relay};
      \begin{align}
      &\mu_{z^{(b_0)}}=\frac{1}{2}{\rm ln}\bigg(\sum_{j=1}^N\sum_{i=1}^M\tau^{(b_0)}_{i,j}\bigg)-\sigma^2_{z^{(b_0)}},\\
       & \sigma^2_{z^{(b_0)}}=\frac{1}{4}{\rm ln}\left(1+\frac{\sum_{j=1}^N\sum_{i=1}^M\left(\tau^{(b_0)}_{i,j}\right)^2\left(e^{4\sigma^2_{X_{ij}}}-1\right)}{\left(\sum_{j=1}^N\sum_{i=1}^M\tau^{(b_0)}_{i,j}\right)^2}\right).
       \end{align}

By means of the above approximation, $P_{be|{b_k}}$ can be evaluated through two-dimensional integral of
    $P_{be|{b_k}}\approx\int_{0}^{\infty}\int_{0}^{\infty}{P_{be|{b_k,\vartheta^{(0)},\vartheta^{(1)}}}}{f(\vartheta^{(0)},\vartheta^{(1)})}d\vartheta^{(0)}d\vartheta^{(1)}$, where ${P_{be|{b_k,\vartheta^{(0)},\vartheta^{(1)}}}}$ is the system BER conditioned on $b_k$, $\vartheta^{(0)}$ and $\vartheta^{(1)}$, and (when Gaussian approximation is used) can be obtained as;
     \begin{align}
   & {P_{be|{b_k,\vartheta^{(0)},\vartheta^{(1)}}}}\!\approx\!Q\!\left(\!\frac{{\tilde{m}}_{\rm MIMO}^{(1)}-{\tilde{m}}_{\rm MIMO}^{(0)}}{\sqrt{{\tilde{m}}_{\rm MIMO}^{(1)}\!+\!{N\sigma }^2_{th}}\!+\!\sqrt{{\tilde{m}}_{\rm MIMO}^{(0)}\!+\!{N\sigma }^2_{th}}}\!\right).
 \end{align}
Note that in the MISO scheme all of the transmitters are pointed to a single receiver; therefore, all of the links have the same fading-free impulse response and channel memory as $h_{0,{\rm {MISO}}}(t)$ and $L_{{\rm {MISO}}}$, respectively. Consequently, when all transmitters have identical {transmitted power} of $P/M$, all links have equal $m^{(s)}_{i,1}$ and $m^{(I,k)}_{i,1}$ as $m^{(s)}_{{\rm {MISO}}}$ and $m^{(I,k)}_{{\rm {MISO}}}$, respectively. Hence, we can rewrite \eqref{M-gaus:4} as $m^{(b_0)}_{\rm MISO}\!=\!m_{\rm MISO}^{(bd)}\!\!+\left(b_0m^{(s)}_{{\rm {MISO}}}+\sum_{k=-L_{{\rm {MISO}}}}^{-1}b_k{m_{{\rm {MISO}}}^{(I,k)}}\right)\varphi^{(M)}$, where $\varphi^{(M)}=\sum_{i=1}^{M}{\alpha^2_{i1}}$, i.e., the sum of $M$ lognormal RVs. As a result, $P_{be|b_k}$ for MISO scheme can be evaluated through one-dimensional integral of $P_{be|{b_k}}\approx\int_{0}^{\infty}{P_{be|{b_k},\varphi^{(M)}}}{f(\varphi^{(M)})}d\varphi^{(M)}$.
Then, if the channel memory is $L_{\rm {max}}$ bits, $P_{be}$ can be obtained by averaging as $P_{be}\!=\!\frac{1}{2^{L_{\rm {max}}}}\!\sum_{b_k}\!\!\!{P_{be|b_k}}$. Note that as ISI increases, this averaging demands more computational time and evaluation of the upper bound BER becomes more advantageous.

 From \eqref{M-gaus:4}, one can observe the destructive effect of ISI on the BER. In other words, experiencing more time spreading in $h_{0,ij}(t)$ or equivalently ${\Gamma }_{i,j}(t)$ increases  $\sum_{k\neq 0}{m_{i,j}^{(I,k)}}$ and decreases $m_{i,j}^{(s)}$. Thereby, it causes an increase in $m^{(0)}$ and a decrease in $m^{(1)}$ which results into larger BERs. Constructive effect of spatial diversity is appeared as combining the fading coefficients of different links which can be approximated as a single lognormal RV with roughly a scaled log-amplitude variance by the number of links \cite{navidpour2007ber}.
%
%
\section{Numerical Results} 
\begin{figure*}[t]
       \centering
       \includegraphics[trim=0cm 0.1cm 0cm 0cm,width=7in,clip]{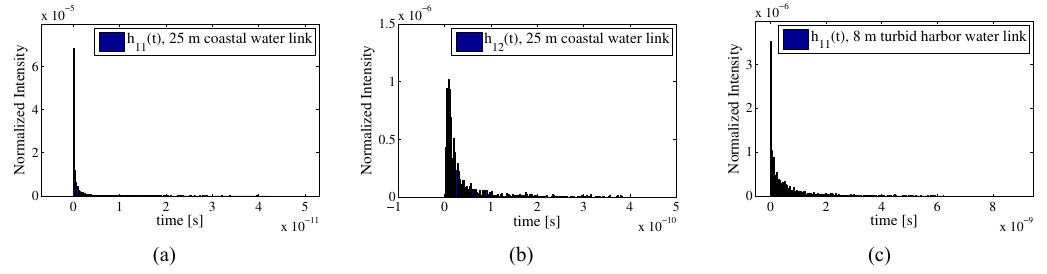}
       \caption{Fading-free impulse responses for a $1\times 2$ SIMO UWOC system in different water types. (a) $h_{0,11}(t)$ of a $25$ {m} coastal water link; (b) $h_{0,12}(t)$ of a $25$ {m} coastal water link; (c) $h_{0,11}(t)$ of an $8$ {m} turbid harbor water link.}
       \label{Fig2_new}
       \end{figure*}
 In this section, we present the numerical results for the BER performance of UWOC systems in various scenarios. We consider lognormal distribution for the channel fading statistics, equal power as $P/M$ for all transmitters, the same fading statistics (log-amplitude variance) for all links and the same aperture area of $A/N$ for all of the receivers, where $A$ is the total aperture area.
In simulating the turbulence-free impulse response by MC method, { we consider coastal water and also turbid harbor water links with absorption and scattering coefficients of $(a,b)=({0.179},{0.219})$ $\rm{m^{-1}}$ and $({0.366},{1.824})$ $\rm{m^{-1}}$, respectively \cite{mobley1994light}}. Other important parameters for MC simulations are listed in Table I.
%
  \begin{table}
    \centering
  \caption{Some of the important parameters used for noise characterization and MC-based channel simulation.}
    \begin{tabular}{||p{2.2in}||p{0.8in}||} \hline
     Coefficient & Value\\ [0.5ex] 
     \hline \hline
     Quantum efficiency, $\eta $ & $0.8$ \\ \hline 
              Optical filter bandwidth, $\triangle \lambda $ & $10$ {nm} \\ \hline 
              Optical filter transmissivity, $T_F$ & $0.8$ \\ \hline 
              Equivalent temperature, $T_e$ & $290$ {K} \\ \hline 
              Load resistance, $R_L$ & $100~\Omega $ \\ \hline 
              Dark current, $I_{dc}$ & $1.226\times {10}^{-9}$ {A}\\
              \hline
     Receiver half angle FOV, $\theta_{\rm FOV}$ & ${40}^0$ \\ \hline 
       MISO schemes aperture diameter, $D^{(\rm MISO)}_0$ & $20$ {cm} \\ \hline 
       Source wavelength, $\lambda $ & $532$ {nm} \\ \hline 
       Water refractive index, $n$ & $1.331$ \\ \hline 
       Source full beam divergence angle, $\theta_{div}$ & $0.02^0$ \\ \hline 
       Photon weight threshold at the receiver, $w_{th}$ & ${10}^{-6}$ \\ \hline 
       Separation distance between the transmitters and between the receiving apertures, $l_0$ & $25$ {cm} \\ \hline
    \end{tabular}
    \vspace{-0.1in}
    \end{table}
    In addition, some of the important parameters for characterization of noises are addressed in this table and the other parameters are exactly the same as those mentioned in \cite{jaruwatanadilok2008underwater,giles2005underwater}. Based on these parameters, noise characteristics are as $n_b\approx {1.8094\times {10}^8}$ $\rm{s^{-1}}$ in $30$ meters deep coastal water, $n_d\approx {76.625\times {10}^8}$ $\rm{s^{-1}}$, and $\sigma^2_{th}/{T_b}=3.12\times10^{15}$ $\rm{s^{-1}}$. Hence, background radiation has a negligible effect on the system performance.
    
{Moreover, in this section we obtain the BER values using numerical simulations to verify the accuracy of our derived expressions. To do so, we generate $10^{7}$ Bernoulli RVs with PDF of $\Pr(b_0)=\frac{1}{2}\delta(b_0)+\frac{1}{2}\delta(b_0-1)$, where $\delta(.)$ is Dirac delta function. Subsequently, $10^7$ pulses of shape $b_0P_i(t)$ will be transmitted to the channel. Each of these $10^7$ pulses will be convolved with the channel fading-free impulse response and multiplied by a lognormal RV to construct $b_0\alpha_{ij}^2P_i(t)*h_{0,ij}(t)=b_0\alpha_{ij}^2\Gamma_{i,j}(t)$. Additionally, a contribution of channel ISI reaches the receiver as described in Eqs. \eqref{eq4} and \eqref{Eq.III-B1}. At the receiver side, the integrated current will be added by a Gaussian distributed noise and the result will be compared with an appropriate threshold (as comprehensively studied in Section III) to detect the received signal. Finally, comparing the transmitted data sequence with the detected one determines the BER value in a specified amount of the transmitted power.}
 
 {In order to see how water turbidity affects the channel loss and temporal dispersion, and also to investigate the channel spatial beam spread, we use MC method to simulate the channel fading-free impulse responses for a $1\times 2$ SIMO transmission in a $25$ {m} coastal water link and also an $8$ {m} turbid harbor water link. The transmitter is pointed to the first receiver and the link between the transmitter and the first receiver has the impulse response of $h_{0,11}(t)$. The second receiver is located in $25$ {cm} center-to-center distance from the first aperture and receives those photons that are reached with much more scattering; with impulse response of $h_{0,12}(t)$. Each receiving aperture has a diameter of $20/\sqrt{2}$ {cm}. Fig. 2 illustrates the simulation results. Comparing Figs. 2(a) and 2(c) shows that while the direct link in a $25$ {m} coastal water has a negligible scattering, an $8$ {m} turbid harbor water link remarkably scatters propagating photons and induces much more delay spread on $h_{0,11}(t)$. In other words, as the water turbidity increases both the channel loss and delay spread considerably raise. As a result, an $8$ {m} turbid harbor water link has a poor channel condition even than a $25$ {m} coastal water link. Moreover, comparing Figs. 2(a) and 2(b) demonstrates that the second receiver mainly receives those photons that have experienced more scattering than the received photons by the first aperture. Accordingly, the channel delay spread in $h_{0,12}(t)$ is by far more than $h_{0,11}(t)$.}

\begin{figure}
     \centering
     \includegraphics[width=3.4in]{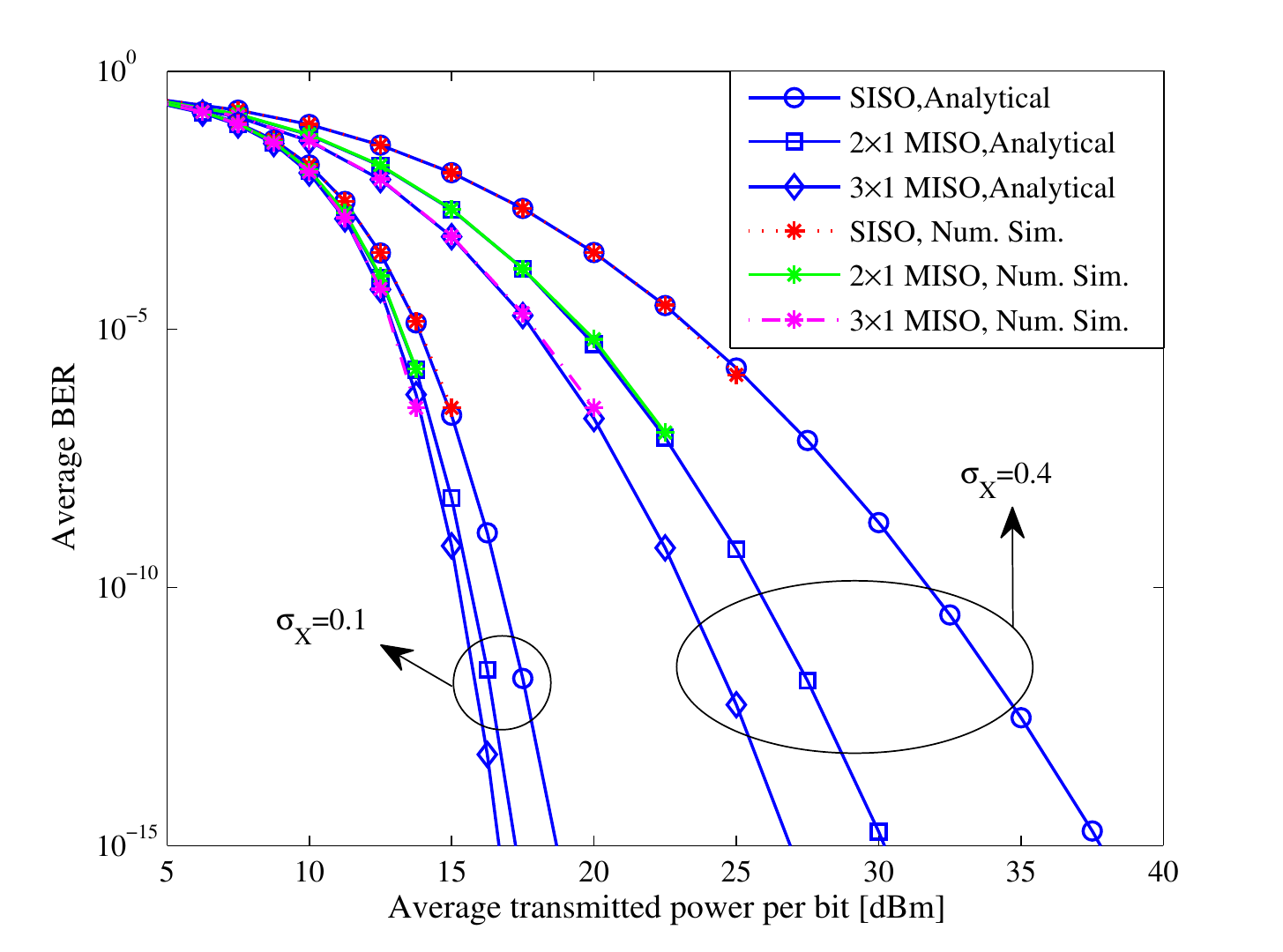}
     \caption{Exact BER of a $25$ {m} coastal water link with SISO, $2\times 1$ MISO and $3\times 1$ MISO configurations, obtained using both analytical expressions and numerical simulations. $R_b=1$ {Gbps}, ${\sigma }_X=0.1$ and $0.4$.}
     \label{Fig2}
     \vspace{-0.1in}
     \end{figure}

 Fig. \ref{Fig2} depicts the exact BER of a $25$ {m} coastal water link with transmitter diversity and data transmission rate of $R_b=1$ {Gbps}. 
This figure also indicates {an excellent match} between the results of analytical expressions and numerical simulations.
  {Here,} we assume that the fading of each link is independent from the others. As it is obvious, increasing the number of independent links provides significant performance improvement in the case of ${\sigma }_X=0.4$, e.g., one can achieve approximately $6$ {dB} and $9$ {dB} performance improvement at the BER of ${10}^{-12}$, using two and three transmitters, respectively. But this benefit relatively vanishes in very weak fading conditions, e.g., ${\sigma }_X=0.1$. This is reasonable, since in very weak turbulence conditions fading has a minuscule effect on the performance but scattering and absorption have yet substantial effects. Hence, in such scenarios multiple transmitters scheme, which combats with impairing effects of fading, does not provide {a} notable performance improvement.
  
 \begin{figure}
       \centering
       \includegraphics[width=3.4in]{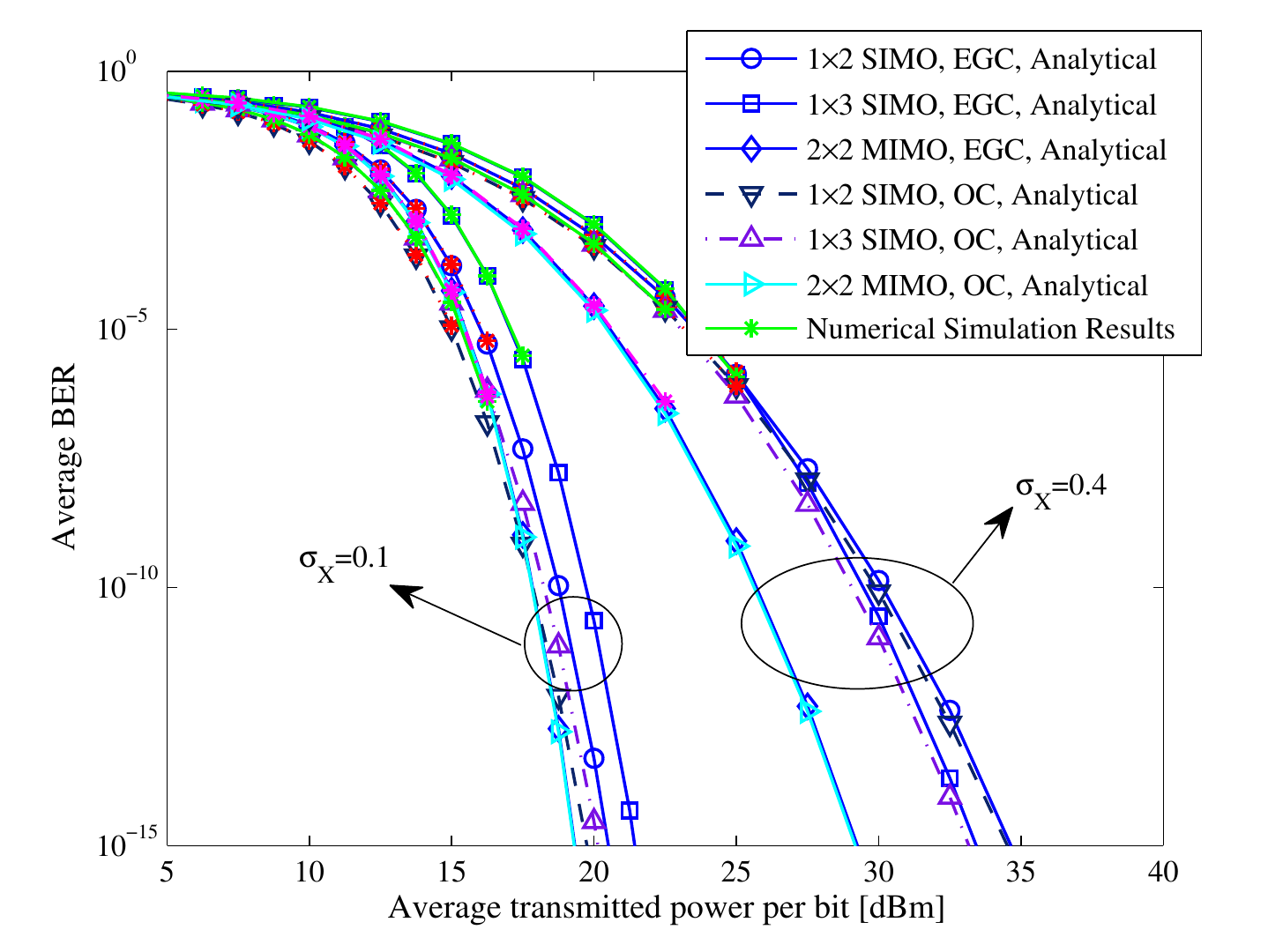}
       \caption{Exact BER of a $25$ {m} coastal water link with $1\times 2$ SIMO, $1\times 3$ SIMO and $2\times 2$ MIMO configurations and optimal/equal gain combiner, obtained using both analytical expressions and numerical simulations. $R_b=1$ {Gbps}, ${\sigma }_X=0.1$ and $0.4$.}
            \label{Fig3}
            \vspace{-0.1in}
       \end{figure}
 
  In Fig. \ref{Fig3}, we assume the same parameters as in Fig. \ref{Fig2} and use our derived analytical expressions to evaluate the exact BER of $1\times 2$ SIMO, $1\times 3$ SIMO and $2\times 2$ MIMO configurations with optimal/equal gain combiner. Comparison between the results {shows} that the performance of EGC is very close to the performance of OC receiver. Therefore, due to its lower complexity, receiver with EGC is more practically interesting. Furthermore, {the good match} between the analytical results and numerical simulations confirms the accuracy of our derived analytical expressions for the system BER.
{Comparing the results of BER for SISO and SIMO schemes demonstrates that a $1\times N'$ SIMO scheme provides better performance than a $1\times N$ SIMO configuration ($N'>N\geq 1)$ only at high signal-to-noise ratios (SNRs) or equivalently low BERs, where fading has more impairing effect than absorption and ISI. This is reasonable, since each receiver in a $1\times N'$ SIMO scheme has $N'/N$ times less aperture area than a $1\times N$ SIMO configuration and also an $N'$-receiver scheme imposes $N'/N$ times more dark current and thermal noise. In low SNRs, absorption and scattering as well as noise have more dominant effects on the BER than fading; therefore, in low SNRs the $1\times N$ SIMO scheme yields better performance than the $1\times N'$ SIMO structure. However, when the channel suffers from relatively notable turbulence, the $1\times N'$ SIMO structure which has more links can better mitigate fading and can compensate for the loss due to the smaller aperture size and excess noise and therefore can yield better performance at higher SNRs. Needless to say that a $2\times 2$ MIMO structure has the same aperture size as a $1\times 2$ SIMO structure and since benefits from more independent links can yield better performance, than a $1\times N'$ SIMO, in all ranges of SNR.}
   One can expect that in a very weak turbulence scenario, such as ${\sigma }_X=0.1$, dividing the receiver aperture to extend the number of independent links can degrade the performance.

\begin{figure}
      \centering
      \includegraphics[width=3.4in]{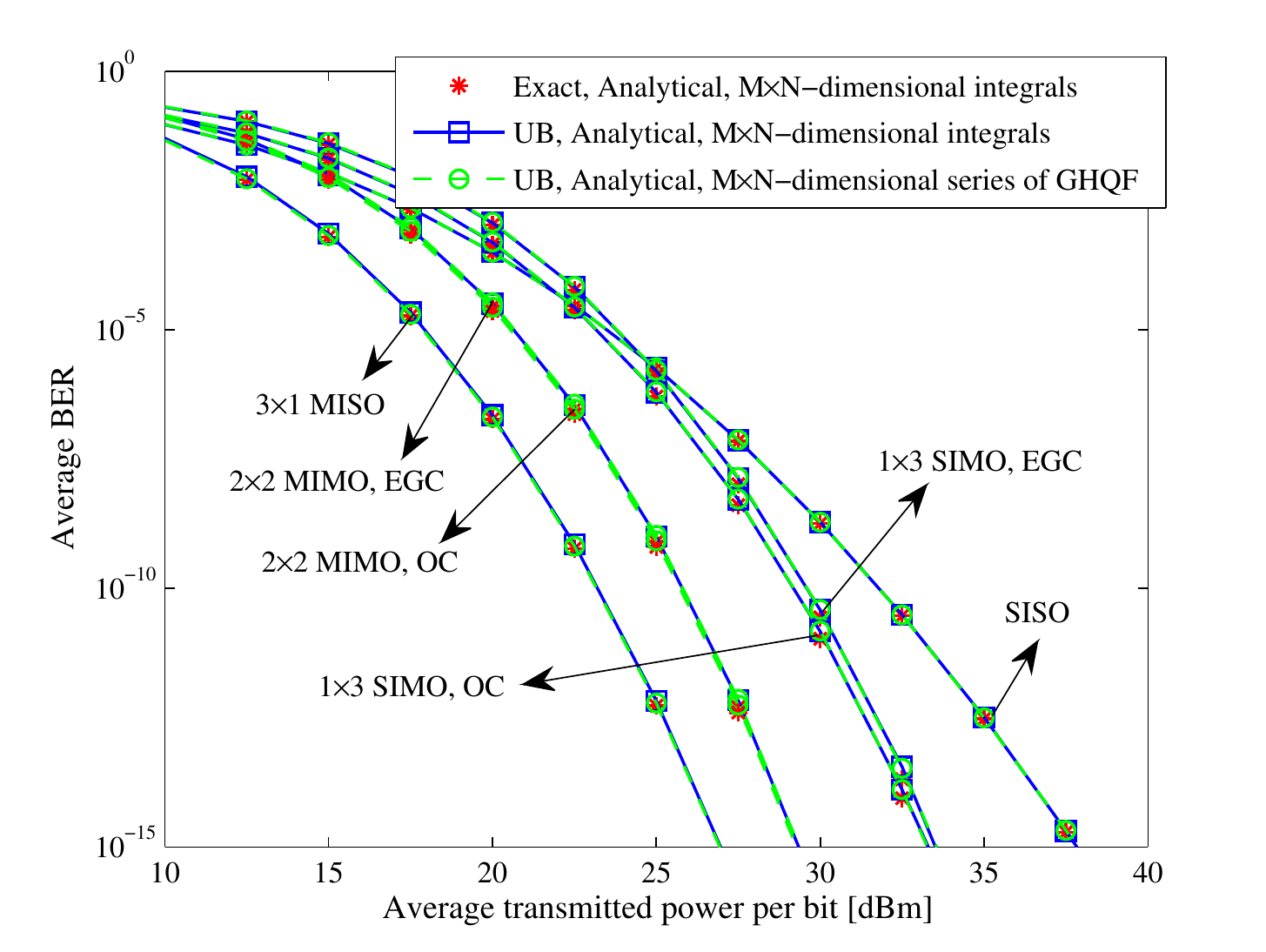}
      \caption{Comparison between the exact and the upper bound BERs of a $25$ {m} coastal water link with  $R_b=1$ {Gbps}, ${\sigma }_X=0.4$, and various configurations. Also the upper bound BERs are calculated with ($M\times N$)-dimensional series, using Gauss-Hermite quadrature formula (GHQF).}
           \label{Fig4}
           \vspace{-0.1in}
      \end{figure}
In Fig. \ref{Fig4} we applied our derived analytical expressions to evaluate the exact and upper bound BERs of a $25$ {m} coastal water link with $R_b=1$ {Gbps} and ${\sigma }_X=0.4$, using ($M\times N$)-dimensional integrals. As it can be seen, the upper bound BER curves have good tightness with the exact BER curves. Therefore, the upper bound BER evaluation can be more preferable since the exact BER calculation may need excessive time for averaging over $b_k$s. Furthermore, the upper bound BERs of various configurations are calculated by ($M\times N$)-dimensional series, using Gauss-Hermite quadrature formula (GHQF). The order of approximation $U$ is assumed to be the same for all of the links, i.e., $U_{ij}=30$. {The excellent match} between the results of GHQF and numerical ($M\times N$)-dimensional integrals demonstrates the usefulness of GHQF in effective calculation of the system BER.

 \begin{figure}
       \centering
       \includegraphics[width=3.4in]{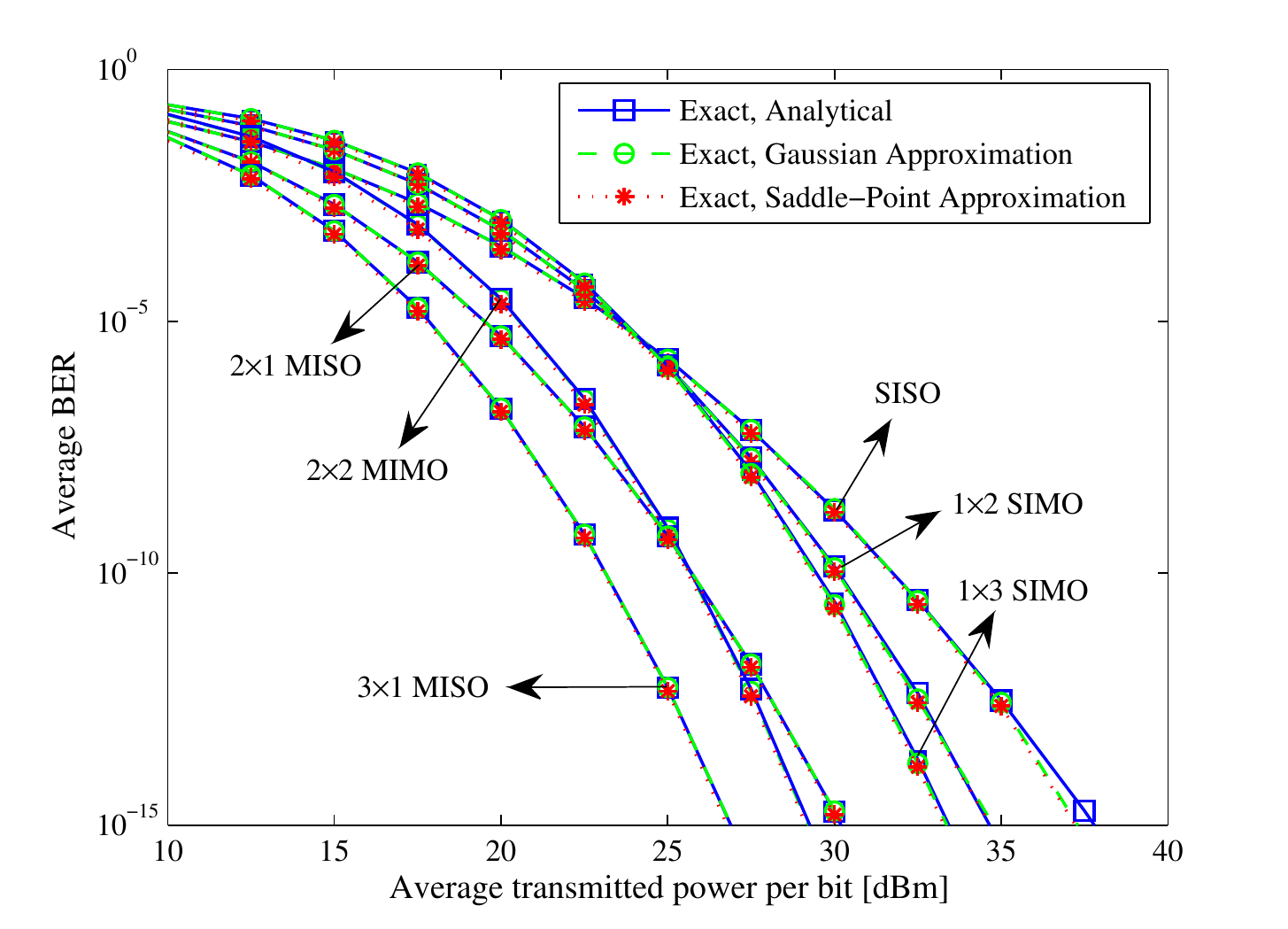}
       \caption{Comparing Gaussian and saddle-point approximations in evaluating the exact BER of a $25$ {m} coastal water link with  $R_b=1$ {Gbps}, ${\sigma }_X=0.4$, and various configurations. Also the results of photon-counting methods are compared with the results of our derived analytical expressions for the system exact BER.}
            \label{Fig5}
            \vspace{-0.1in}
       \end{figure}
Fig. \ref{Fig5} compares Gaussian and saddle-point approximations in evaluating the exact BER of a $25$ {m} coastal water link with  $R_b=1$ {Gbps}, ${\sigma }_X=0.4$, and various configurations. It is observed that Gaussian approximation can provide relatively the same results as saddle-point approximation. Therefore, due to its simplicity and acceptable accuracy, Gaussian approximation can be considered as a reliable photon-counting method for the system BER evaluation. Moreover, the results of our derived analytical expressions are compared with those of photon-counting methods. {The good match} between the results of analytical expressions and photon-counting methods further confirms the validity of our assumption in neglecting the signal-dependent shot noise in our analytical derivations.
{
\begin{table*}[t]
  \centering
  \caption{The computation time ($\tau_{\rm comp}$) and BER values of different methods for various system configurations. Average transmitted power per bit $=20$ {dBm}, $R_b=1$ {Gbps}, $\sigma_X=0.4$, and $d_0=25$ {m} coastal water.}
  \label{T3}
\begin{tabular}{|p{2.3cm}|p{0.65cm}||p{1.6cm}|p{1.6cm}|p{1.6cm}|p{1.6cm}|p{1.6cm}|p{1.6cm}|}
\hline
\multicolumn{2}{|l||}{\backslashbox{Method}{Configuration}}
&{SISO}&{$2\times 1$ MISO}&{$1\times 2$ SIMO with OC}&{$1\times 2$ SIMO with EGC}&{$2\times 2$ MIMO with OC}&{$2\times 2$ MIMO with EGC}\\ \hline\hline

\multirow{2}{*}{Analytical} & {$\tau_{\rm comp}$} &$0.05750$ {s}&$1.5257$ {s}&$1.7164$ {s}&$1.6162$ {s}&$140.6670$ {s}&$127.8600$ {s} \\ \cline{2-8} 
                  & $P_{be}$ &$\!\!3.050\!\times\!10^{-4}\!$&$\!\!5.088\!\times\!10^{-6}\!$&$\!\!3.916\!\times\!10^{-4}\!$&$\!\!6.408\!\times\!10^{-4}\!$&$\!\!2.188\!\times\!10^{-5}\!$&$\!\!2.639\!\times\!10^{-5}\!$\\
\hline \hline
\multirow{2}{*}{Gaussian approx.} & {$\tau_{\rm comp}$} &$0.05717$ {s}&$0.8511$ {s}&$-----$&$0.8668$ {s}&$-----$&$116.3134$ {s}\\ \cline{2-8} 
                  & $P_{be}$ &$\!\!3.056\!\times\!10^{-4}\!$ &$\!\!5.106\!\times\!10^{-6}\!$&$-----$&$\!\!6.372\!\times\!10^{-4}\!$&$-----$&$\!\!2.607\!\times\!10^{-5}\!$\\
\hline \hline
\multirow{2}{*}{Saddle-point approx.} & {$\tau_{\rm comp}$} &$38.4672$ {s}&$1969.026$ {s}&$-----$&$7799.913$ {s}&$-----$&$432614.65$ {s} \\ \cline{2-8} 
                  & $P_{be}$ &$\!\!3.004\!\times\!10^{-4}\!$&$\!\!5.139\!\times\!10^{-6}\!$&$-----$&$\!\!6.288\!\times\!10^{-4}\!$&$-----$&$\!\!2.639\!\times\!10^{-5}\!$\\
\hline
\end{tabular}
\end{table*}
}

{ After confirming the accuracy of our derived analytical expressions, a comparison between the computation time of different methods would be interesting. Table II shows the computation time and BER values of different methods for various system configurations in a $25$ {m} coastal water link with $\sigma_X=0.4$, $R_b=1$ {Gbps}, and average transmitted power per bit of $20$ {dBm}. The values in this table are obtained through $(M\times N)$-dimensional integration over fading coefficients as well as averaging over all $2^{L_{\rm max}}$ sequences for $b_k$s. In order to perform the integration, we have taken $N_{\rm samp}$ equidistant samples from the interval $\alpha_{ij}\in[0.0001,5]$. Here, we have chosen $L_{\rm max}=3$, $N_{\rm samp}=30$ for MIMO configuration, and $N_{\rm samp}=100$ for the other schemes. As it can be seen, saddle-point approximation has by order of magnitude larger computation time, since it involves solving some complex nonlinear equations to obtain the conditional BERs. On the other hand, our derived analytical expressions and Gaussian approximation have relatively the same calculation time, since both involve Gaussian-Q function which is a fast-calculable built-in function in Matlab. Moreover, all three methods yield similar results for the system average BER, confirming the accuracy of our derived analytical expressions and the validity of our assumption in Section III, i.e., the negligibility of signal-dependent shot noise. A more advantage of our derived analytical expressions becomes visible when we apply the approximation to the sum of RVs. In this case, as it is elaborated in Section III, in the case of transmitter diversity ($N=1$) and also receiver diversity with EGC, the $(M\times N)$-dimensional averaging integration can be evaluated through an equivalent one-dimensional integral using our derived analytical expressions, while Gaussian approximation only in the case of transmitter diversity leads to a one-dimensional integral, as shown in Section IV.} {Accordingly, when approximation to the sum of RVs is applied, our derived analytical expressions can more quickly predict the system performance. Therefore, our analytical approach is more advantageous and convenient from the computation time and ease of mathematical manipulation points of view; and hence is recommended to be employed for analysis of MIMO UWOC systems.}
%

\begin{figure}
         \centering
         \includegraphics[width=3.4in]{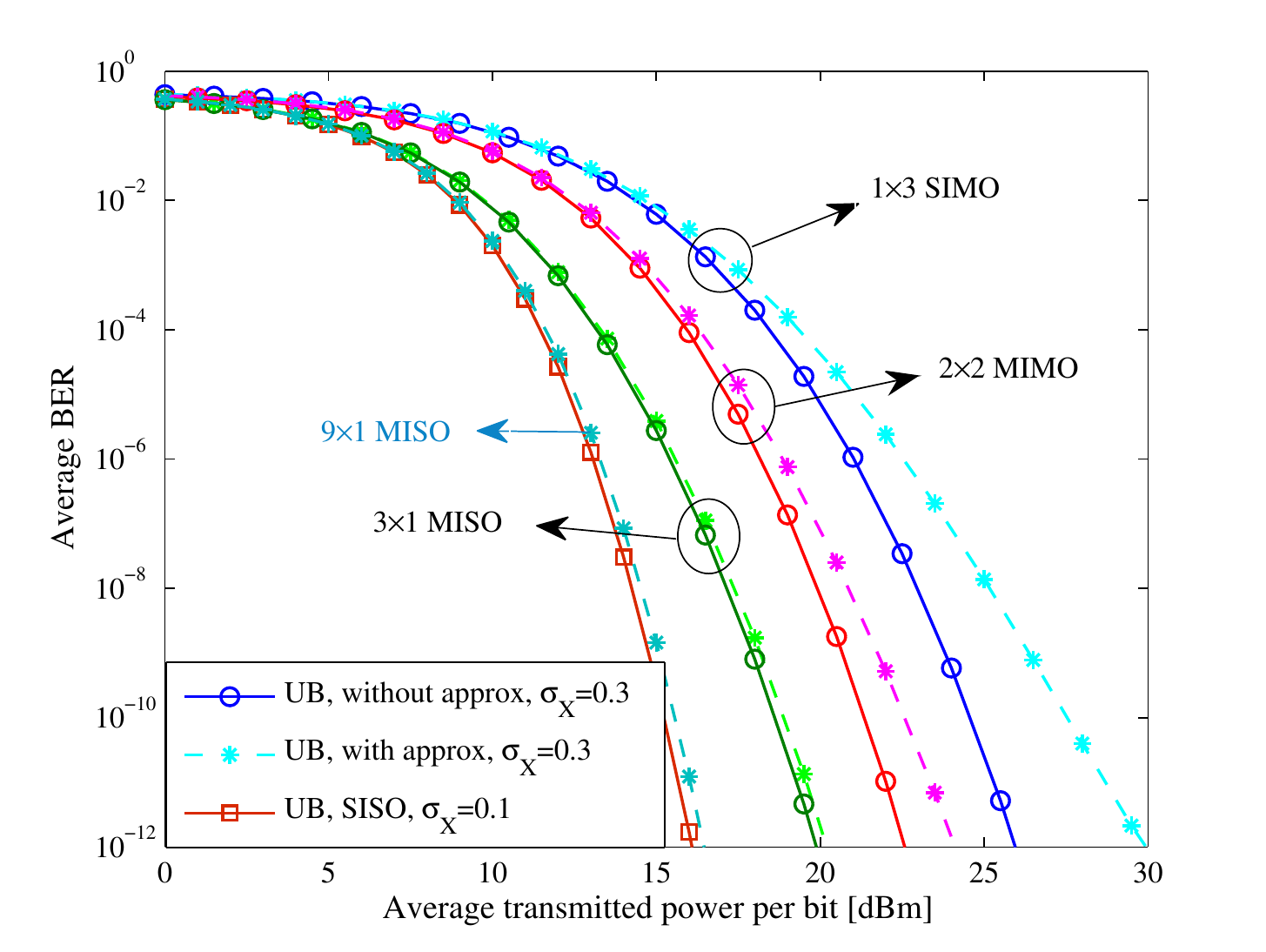}
         \caption{Upper bound and approximated upper bound on the BER of a $25$ {m} coastal water link with different configurations, obtained using Gaussian approximation. $R_b=0.5$ {Gbps}, ${\sigma }_X=0.3$ and $0.1$.}
         \label{Fig6}
         \vspace{-0.1in}
\end{figure}

In Fig. \ref{Fig6}, the BER performance of a $25$ {m} coastal water link with $R_b=0.5$ {Gbps} and $\sigma_X=0.3$ is depicted for different configurations.
  Also the sum of independent lognormal RVs in \eqref{M-gaus:4} is approximated with a single lognormal RV and the BER is evaluated through the approximated one-or two-dimensional integrals. As it can be seen, relatively good match exists between the results of the approximated one-or two-dimensional integrals and the exact $(M\times N)$-dimensional integral of $P_{be}=\int_{\vec{\boldsymbol{\alpha}}}P_{be|\vec{\boldsymbol{\alpha}}}{f_{\vec{\boldsymbol{\alpha}}}\left(\vec{\boldsymbol{\alpha}}\right)}d\vec{\boldsymbol{\alpha}}$. However, the discrepancy increases when receiver diversity is used. Moreover, the BER performance of a SISO link with ${\sigma }_X=0.1$ is compared with the BER curve of a $9\times1$ MISO link with ${\sigma }_X=0.3$, and approximately the same result is observed; therefore, spatial diversity manifests its effect as a reduction in the fading log-amplitude variance.

\begin{figure}
        \centering
        \includegraphics[width=3.4in]{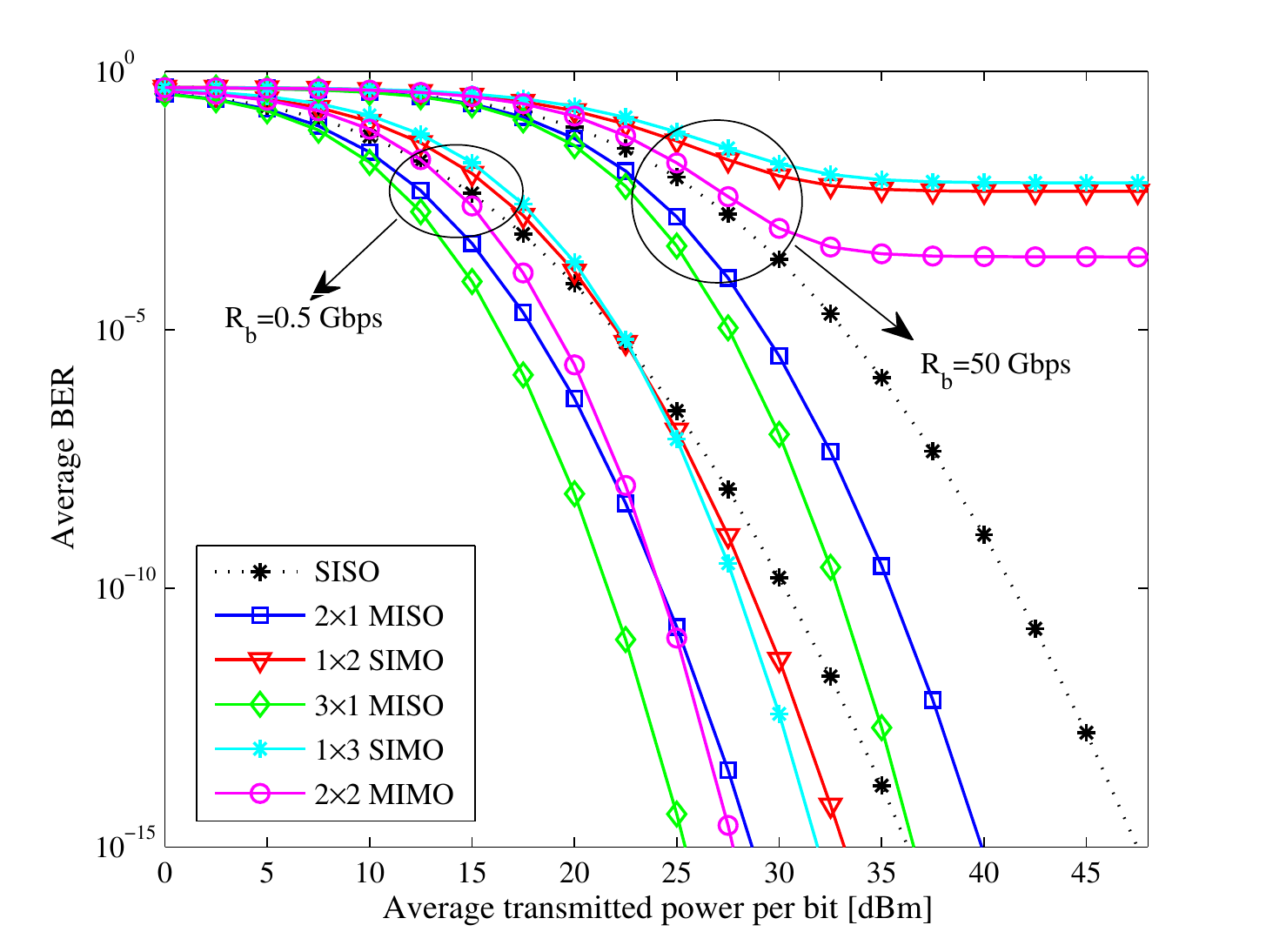}
        \caption{Effect of ISI on the performance of a $25$ {m} coastal water link with different configurations. ${\sigma }_X=0.4$, $R_b= 0.5$ {Gbps} and $50$ {Gbps}.}
        \label{Fig7}
        \vspace{-0.1in}
        \end{figure}
{In order to investigate how ISI affects the performance of different configurations, in Fig. 8 the upper bound on the BER of a $25$ {m} coastal water link with ${\sigma }_X=0.4$ and various configurations is illustrated for a typical data rate of $R_b=0.5$ {Gbps} and also an extremely large data rate of $R_b=50$ {Gbps}. As it can be seen, increasing the transmission data rate considerably degrades the performance of all configurations, especially for multiple receivers schemes where a lower bound on the BER is observable. This behavior can thoroughly be inferred from Figs. 2(a) and 2(b). In other words, in MISO schemes where a single receiver is employed, all of the transmitters are pointed to a single receiver; hence, all of the links from each transmitter to the receiver have impulse response of $h_{0,11}(t)$, shown in Fig. 2(a). In fact, based on Fig. 2(a), in both of the used bit duration times, i.e., $T_b=2\times10^{-9}$ {s} and $2\times10^{-11}$ {s}, the direct link delay spread is not much more than the bit duration time to see a saturation behavior in the BER of MISO schemes. Nevertheless, as the bit duration time decreases (data rate increases) the destructive effect of ISI becomes more severe and performance of both SISO and MISO schemes degrades; however, MISO transmission in both of the aformentioned data rates can yet yield a better performance than SISO transmission. On the other hand, in SIMO and MIMO schemes where more than one receiver is used, each transmitter is pointed to one of the receivers; however, due to the multiple scattering of UWOC channels some of the transmitted photons by each transmitter may reach the other receiving apertures that are not necessarily pointed to that transmitter. In other words, in SIMO and MIMO schemes, there always exist photons that have reached the receiver plane through indirect path, i.e., with the indirect link impulse response, such as $h_{0,12}(t)$ shown in Fig. 2(b). As it has been shown in Figs. 2(a) and 2(b), for $R_b=0.5$ {Gbps} delay spread of both the direct and indirect links is enough smaller than the bit duration time; hence, ISI does not considerably degrade the performance of SIMO and MIMO schemes with $R_b=0.5$ {Gbps}. Nevertheless, for $R_b=50$ {Gbps}, although the direct link temporal spread is yet enough smaller than the bit duration time, the indirect link delay spread is much more than the bit duration time. Therefore, for the case of $R_b=50$ {Gbps} the destructive contribution of indirect link extremely increases ISI for multiple receivers schemes, where even increasing the transmission power cannot compensate for this degradation. In other words, in very large data rates the destructive effect of indirect link saturates the BER of multiple receivers schemes to a lower bound. It should be noted that decreasing the separation distance between the receiving apertures alleviates this time spreading, but this approach increases the spatial correlation between different links' signals, which degrades the system performance (as it is shown in Fig. 11). In addition, we can reduce the ISI effect by decreasing the receiver aperture size, but this scheme induces more loss on the received optical signal.}
%
\begin{figure}
        \centering
        \includegraphics[width=3.4in]{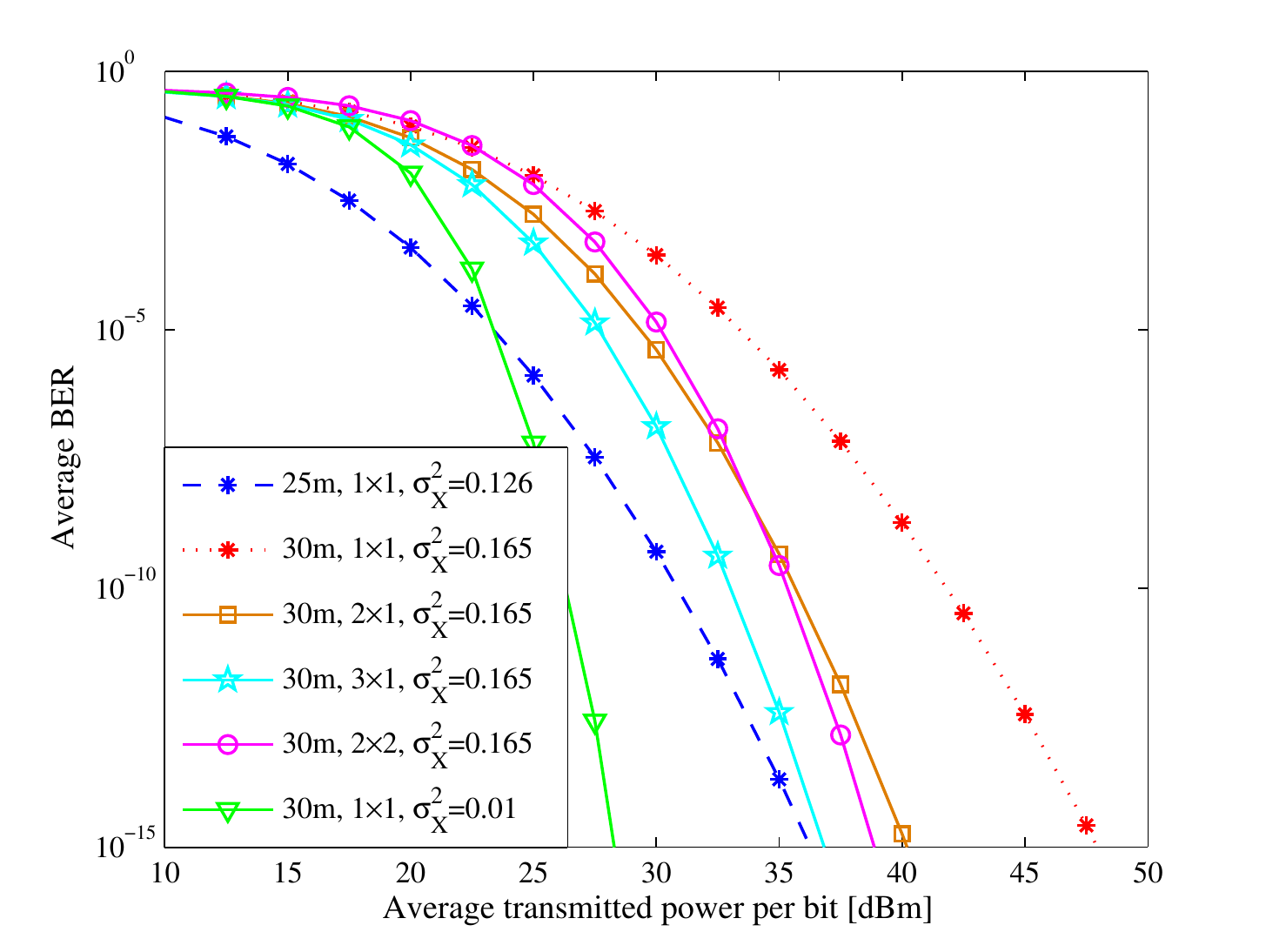}
        \caption{Comparison between the performance of a $30$ {m} coastal water link with different configurations and a $25$ {m} SISO link, both operating at $R_b=2$ {Gbps}.}
        \label{Fig9}
        \vspace{-0.1in}
 \end{figure}     
   
 {Fig. \ref{Fig9} demonstrates the ability of spatial diversity in extending the viable communication range of UWOC systems, as one of the important objectives of this paper.} In this figure we want to show that although increasing the communication distance increases absorption (loss), scattering (ISI) and turbulence (fading), yet using spatial diversity one can achieve better performance than SISO links with smaller distances and therefore can increase the viable communication range. To show that, we consider a coastal water with scintillation parameters of $\varepsilon ={{10}^{-5}}$ $\rm {m^2/{s^3}}$, $w=-3$ and $\ {\chi }_T={4\times {10}^{-7}}$ $\rm{K^2/{s}}$, and numerically evaluate Eq. \eqref{SI} to find the scintillation index of a plane wave for different link ranges. Based on our numerical results we find that for $d_0=25$ {m} and $30$ {m} the
 log-amplitude variance ${\sigma }^2_X$ is $0.126$ and $0.165$, respectively. As it is obvious in Fig. 9, only $5$ {m} ($\%20$) increase on the communication range remarkably degrades the system performance, e.g.,  approximately $12$ {dB} degradation is observed at the BER of ${10}^{-12}$. But as it can be seen, increasing the number of independent links or equivalently mitigating fading deteriorations considerably improves the system performance. Since spatial diversity manifests itself as a reduction in fading variance \cite{navidpour2007ber}, we can conclude that there exists a configuration with spatial diversity at link range of $30$ {m} which performs similar to a SISO link at that range but with less fading variance, e.g., ${\sigma }^2_X=0.01$. Hence, in this figure the performance of a $30$ {m} SISO link with ${\sigma }^2_X=0.01$ is also depicted for the sake of comparison and obviously it can yield better performance than a $25$ {m} SISO link, especially for lower error rates. Therefore, one can achieve better performance even in longer link ranges by employing spatial diversity technique.
%
We should emphasize that spatial diversity can provide more performance enhancement rather than those are presented in this paper; when the channel suffers from strong turbulence \cite{tsiftsis2009optical}. But since this paper is focused on weak oceanic turbulence, we only considered channels with ${\sigma }^2_I<1$ \cite{andrews2001laser}.

\begin{figure}[t]
       \centering
       \includegraphics[trim=0cm 0cm 0cm 0cm,width=3.4in,clip]{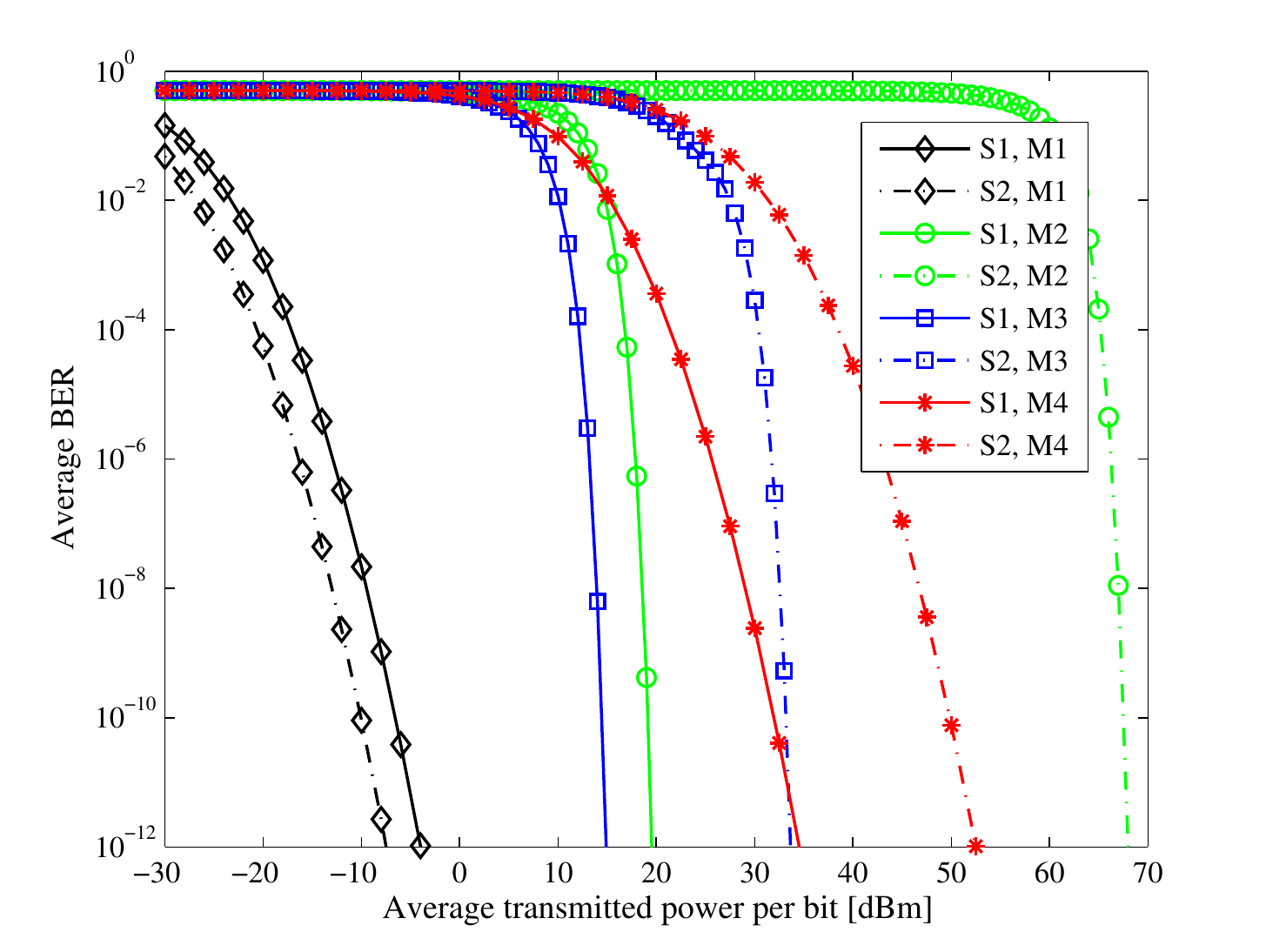}
       \caption{The exact BER of a SISO UWOC link for two separate scenarios (S1: a $25$ {m} coastal water link with $R_b=1$ {Gbps}, and S2: a $10$ {m} turbid harbor water link with $R_b=200$ {Mbps}) and four different models for the channel impulse response described within the text (M1, M2, M3, and M4).}
       \label{new figure}
       \vspace{-0.1in}
       \end{figure}
       
{In order to better observe the effects of different factors on the channel impulse response and hence on the system performance, we consider four different channel models and evaluate the BER performance of a SISO UWOC link for each of the considered channel models. The first model (M1) only considers turbulence effect and ignores absorption and scattering effects, i.e., $h_1(t)=\alpha^2\delta(t-d_0/v)$, where $v$ is the propagation speed of light through water. The second model (M2) only considers absorption and scattering using Beer's law \cite{mobley1994light} and neglects turbulence effect, i.e., $h_2(t)=e^{-cd_0}\delta(t-d_0/v)$. The third model (M3) obtains the channel fading-free impulse response using MC numerical simulations but ignores turbulence effect, i.e., $h_3(t)=h_0(t)$. And the fourth model (M4), similar to the rest of the paper, takes all of the three impairing effects into account while considers absorption and scattering effects based on MC simulations and turbulence effect as a multiplicative fading coefficient, i.e., $h_4(t)=\alpha^2h_0(t)$. We also consider two different scenarios: S1; a $25$ {m} coastal water link with transmission rate $R_b=1$ {Gbps}, and S2; a $10$ {m} turbid harbor water link with $R_b=200$ {Mbps}. All of the system parameters are the same as those are listed in Table I, and the log-amplitude variance for both of the scenarios is considered to be $\sigma^2_X=0.16$. 
As it is shown in Fig. \ref{new figure}, based on the results of the accurate channel model, i.e., the fourth channel model M4, a $25$ {m} coastal water link has approximately $18$ {dB} better performance than a $10$ {m} turbid harbor water link even for $5$ times larger transmission rates. Furthermore, as expected, ignoring the turbulence effect through considering the channel impulse response as the third model M3, underestimates the channel impairments and significantly increases the slope of BER curves. On the other hand, considering the channel impulse response as the second model overestimates the channel attenuation and shifts the BER curves to the right side when compared with the third channel model results. This is because Beer's law ignores those photons which may reach the receiver through multiple scattering. The overestimation of Beer's law for turbid harbor water link is by far more than that of coastal water link due to the higher values of attenuation length defined as $\tau_{atn}=cd_0$ (for the above two scenarios $\tau_{atn, S1}=9.95$ and $\tau_{atn, S2}=21.9$) \cite{tang2014impulse,akhoundi2015cellular}. In other words, in turbid harbor waters the occurance of multiple scattering is very prevalent and many of the transmitted photons may reach the receiver after scattering many times, and ignoring multiple scattering in such links, through use of Beer's law, results into an extreme overestimation of the channel loss. Finally, ignoring absorption and scattering effects and only considering turbulence effects as a multiplicative fading coefficient, through use of the first channel model M1, significantly underestimates the channel degrading effects and considerably shifts the BER curves to the left side. This is mainly because in coastal and turbid harbor UWOC channels with weak turbulence conditions, absorption and scattering have much more impairing effects than turbulence, and ignoring their effects remarkably underestimates the channel impairments.}

\begin{figure}
          \centering
          \includegraphics[width=3.4in]{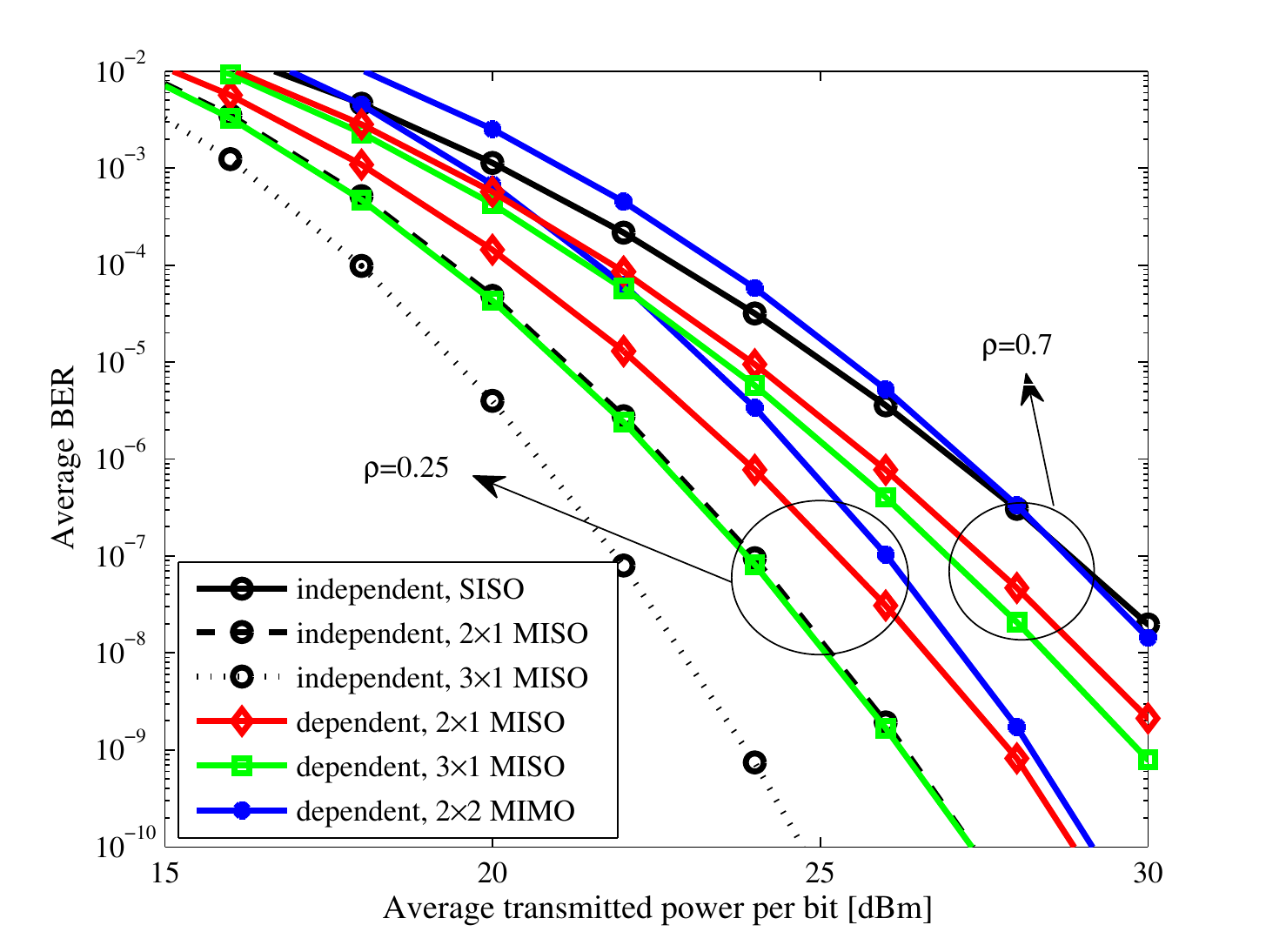}
          \caption{Effect of spatial correlation on the performance of a $25$ {m} coastal water link with ${\sigma }_X=0.4$, $R_b=2$ {Gbps}, $\rho=0.25$ and $0.7$.}
          \label{Fig10}
          \vspace{-0.1in}
   \end{figure}
\indent{As it has been observed, spatial diversity provides a significant performance improvement but under the assumption of independent links which is sometimes practically infeasible \cite{abou2011cooperative}. Therefore, in practice received signals by different links may have correlation. In Fig. \ref{Fig10}, we investigate the effect of spatial correlation on the performance of MIMO UWOC systems in a similar approach to \cite{navidpour2007ber}. Given an $M$ by $M$ correlation matrix at the transmitter side as ${\rm \textbf{R}_T}$ and also an $N$ by $N$ correlation matrix at the receiver side as ${\rm \textbf{R}_R}$, the spatial correlation matrix of the MIMO channel can be obtained by the Kronecker product of the spatial correlation matrices of the transmitter and receiver, i.e., ${\rm \textbf{R}_{MIMO}}={\rm \textbf{R}_T} {\otimes} {\rm \textbf{R}_R}$, which is of size $M\times N$ by $M\times N$ \cite{navidpour2007ber}. Then, given ${\rm \textbf{X}}=[X_1,X_2,...,X_{M\times N}]$ as a vector of $M\times N$ Gaussian distributed RVs with mean zero and variance one, the vector of new RVs, ${\rm \textbf{X}'}=[X'_1,X'_2,...,X'_{M\times N}]$, corresponding to the Gaussian distributed log-amplitude factors with correlation matrix of ${\rm \textbf{R}_{MIMO}}$, can be obtained as;
\begin{align}
{\rm \textbf{X}'}=\sigma_X{\rm \textbf{X}}{\boldmath{\mathcal{C}}}-{\Sigma},
\end{align}
where $\Sigma=[\sigma^2_X,...,\sigma^2_X]$ is a vector of $M\times N$ elements of $\sigma^2_X$, and ${\boldmath{\mathcal{C}}}$ is an $M\times N$ by $M\times N$ upper triangular matrix obtained from the Cholesky decomposition of the correlation matrix ${\rm \textbf{R}_{MIMO}}$ as ${\rm \textbf{R}_{MIMO}}={\boldmath{\mathcal{C}}}^*{\boldmath{\mathcal{C}}}$, where ${\boldmath{\mathcal{C}}}^*$ denotes the complex conjugate transpose of ${\boldmath{\mathcal{C}}}$. Finally, the new correlated lognormal RVs, corresponding to the fading coefficients of different links, can be obtained as $\alpha'^2_{ij}=\exp(2X'_{ij})$, and the average BER can be evaluated from these correlated fading coefficients by numerical calculation of multi-dimensional integrals.}
 In the simulations of Fig. \ref{Fig10}, we have considered a $25$ {m} coastal water link with ${\sigma }_X=0.4$ and $R_b=2$ {Gbps}, and evaluated the upper bound BER of different configurations in several cases, i.e., independent links and correlated links with correlation values of $b\left(l_0\right)=\rho =0.25$ and $0.7$. Comparing the results shows that the performance loss is much more severe for larger correlation values, e.g., $2$ {dB} and $6$ {dB} degradation can be observed in the performance of a $3\times 1$ MISO link with the BER of $10^{-10}$ and with $\rho =0.25$ and $0.7$, respectively. Also a $3\times 1$ MISO link with $\rho =0.25$ yields the same performance as an independent $2\times 1$ MISO link, i.e., {the} diversity order is decreased by one. Therefore, the performance enhancement of spatial diversity in highly correlated weak turbulent channels may be insignificant.
 \section{Conclusion}
 In this paper, we studied the performance of MIMO UWOC systems with OOK modulation and equal gain or optimal combiner. {Our derivations were based on a general channel modeling which appropriately takes all of the channel impairments into account. In particular, we obtained the channel fading-free impulse response using MC numerical simulations and included turbulence effects as a multiplicative fading coefficient.}
 Closed-form solutions for the system BER expressions obtained in the case of lognormal underwater fading channels, relying on Gauss-Hermite quadrature formula as well as approximation to the sum of lognormal random variables. We also applied photon-counting method to evaluate the system BER in the presence of shot noise. {The excellent match} between the results of analytical expressions and photon-counting method confirmed the validity of our assumptions in derivation of analytical expressions for the system BER. Furthermore, our numerical results indicated that EGC is more practically interesting, due to its lower complexity and its close performance to optimal combiner.
  In addition to evaluating the exact BER, also the upper bound on the system BER has been evaluated and excellent tightness between the exact and upper bound BER curves has been observed. Moreover, {the good match} between the results of numerical simulations and analytical expressions verified the accuracy of our derived analytical expressions for the system BER.
   Our numerical results showed that spatial diversity manifests its effect as a reduction in fading variance and hence can significantly improve the system performance and increase the viable communication range. In particular, a $3\times1$ MISO transmission in a $25$ {m} coastal water link with log-amplitude variance of $0.16$ can introduce $8$ {dB} performance improvement at the BER of $10^{-9}$. We also observed that spatial correlation can impose a severe loss on the performance of MIMO UWOC systems. Specifically, correlation value of $\rho=0.25$ between the links of a $3\times 1$ MISO UWOC system with $\sigma_X=0.4$ decreases the order of diversity by one.
 Finally, we should emphasize that although all {of} the numerical results of this paper are based on lognormal distribution, many of our derivations can be used for any other fading statistical distribution.

\appendices
\section{Decision Rule for MIMO UWOC System with OC}
In this appendix, we obtain the decision rule for BER evaluation of MIMO UWOC system with OC. {Based on the ML detection rule and assuming} the availability of perfect CSI, the symbol-by-symbol receiver which does not have any knowledge to $\{b_k\}_{k=-L_{i,j}}^{-1}$, adopts the following metric for optimum combining \cite{navidpour2007ber};
\begin{align}\label{Eq.III-B2}
\Pr\left(\vec{\boldsymbol{r}}|b_0=1,{\vec{\boldsymbol{\alpha}}}\right){\underset{0}{\overset{1}{\gtrless}}}\Pr\left(\vec{\boldsymbol{r}}|b_0=0,\vec{\boldsymbol{\alpha}}\right),
\end{align}
where $\vec{\boldsymbol{r}}=(r_1,r_2,...,r_N)$ is the vector of different branches' integrated received current and $\vec{\boldsymbol{\alpha}}=(\alpha_{11},\alpha_{12},...,\alpha_{MN})$ is the fading coefficients vector. The conditional probabilities for the ``ON" and ``OFF" states are respectively given as;
\begin{align}\label{Eq.III-B3}
&\Pr\left(\vec{\boldsymbol{r}}|b_0=1,{\vec{\boldsymbol{\alpha}}}\right)=\nonumber\\
&\frac{1}{(2\pi\sigma^2_{T_b})^{N/2}}\exp\left(\frac{-1}{2\sigma^2_{T_b}}\sum_{j=1}^{N}\left[r_j-\sum_{i=1}^{M}\alpha_{ij}^2\gamma^{(s)}_{i,j}\right]^2\right),
\end{align}
\begin{align}\label{Eq.III-B4}
\Pr\left(\vec{\boldsymbol{r}}|b_0=0,{\vec{\boldsymbol{\alpha}}}\right)=\frac{1}{(2\pi\sigma^2_{T_b})^{N/2}}\exp\left(\frac{-1}{2\sigma^2_{T_b}}\sum_{j=1}^{N}r_j^2\right).
\end{align}
Replacing \eqref{Eq.III-B3} and \eqref{Eq.III-B4} in \eqref{Eq.III-B2} and dropping the common terms out, the decision rule simplifies to;
\begin{align}\label{Eq.III-B5}
\sum_{j=1}^{N}2r_j\sum_{i=1}^{M}\alpha_{ij}^2\gamma^{(s)}_{i,j}\underset{0}{\overset{1}{\gtrless}}\sum_{j=1}^{N}\left(\sum_{i=1}^{M}\alpha_{ij}^2\gamma^{(s)}_{i,j}\right)^2.
\end{align}
\section{($M\times N$)-Dimensional Series of Gauss-Hermite Quadrature Formula}
In this appendix, we show how ($M\times N$)-dimensional averaging integrals over fading coefficients can effectively be calculated using Gauss-Hermite quadrature formula [37, Eq. (25.4.46)]. More specifically, we calculate $P^{\rm MIMO}_{be|b_0,b_k}=\int_{\vec{\boldsymbol{\alpha}}}P^{\rm MIMO}_{be|b_0,{\vec{\boldsymbol{\alpha}}},b_k}f_{\vec{\boldsymbol{\alpha}}}(\vec{\boldsymbol{\alpha}})d\vec{\boldsymbol{\alpha}}$ with ($M\times N$)-dimensional series where $P^{\rm MIMO}_{be|b_0,{\vec{\boldsymbol{\alpha}}},b_k}$ is defined in \eqref{Eq.III-B6} and \eqref{Eq.III-B7} (e.g., for OC) for $b_0=1$ and $b_0=0$, respectively. Based on \eqref{eq9} and [37, Eq. (25.4.46)], for any function $g(\alpha^2_{ij})$ averaging over lognormal distributed fading coefficient $\alpha_{ij}$ can be calculated with a finite series as;
\begin{align} \label{app_a_new>>eq1}
&\int_{0}^{\infty}g(\alpha_{ij}^2)f_{\alpha_{ij}}(\alpha_{ij})d\alpha_{ij}\approx\nonumber\\ &\frac{1}{\sqrt{\pi}}\sum_{q_{ij}=1}^{U_{ij}}w^{(ij)}_{q_{ij}}g\Big({\alpha^2_{ij}}= \exp\left(2x^{(ij)}_{q_{ij}}\sqrt{2\sigma^2_{X_{ij}}}+2\mu_{X_{ij}}\right)\Big).
\end{align}
Further, the validity of Eq. \eqref{app_a_new>>eq2}, shown at the top of the next page, can be verified by induction for any function of $M\times N$ lognormal distributed fading coefficients $g(\alpha^2_{11},\alpha^2_{21},...,\alpha^2_{MN})$.
\begin{figure*}[!t]
\normalsize
\setcounter{equation}{43}
\begin{align} \label{app_a_new>>eq2}
&\int_{\alpha_{11}=0}^{\infty}\int_{\alpha_{21}=0}^{\infty}...\int_{\alpha_{MN}=0}^{\infty}
g(\alpha^2_{11},\alpha^2_{21},...,\alpha^2_{MN})f_{\alpha_{11}}(\alpha_{11})f_{\alpha_{21}}(\alpha_{21})...f_{\alpha_{MN}}(\alpha_{MN})
d\alpha_{11}d\alpha_{21}...d\alpha_{MN}\nonumber\\
&\approx \frac{1}{{\pi}^{M\times N/2}}\sum_{q_{11}=1}^{U_{11}}w^{(11)}_{q_{11}}\sum_{q_{21}=1}^{U_{21}}w^{(21)}_{q_{21}}...\sum_{q_{MN}=1}^{U_{MN}}w^{(MN)}_{q_{MN}}
g\Big({\alpha^2_{11}}= \exp\left(2x^{(11)}_{q_{11}}\sqrt{2\sigma^2_{X_{11}}}+2\mu_{X_{11}}\right),\nonumber\\
&~~~~~~~~~~~~~~~~~~~~~~{\alpha^2_{21}}= \exp\left(2x^{(21)}_{q_{21}}\sqrt{2\sigma^2_{X_{21}}}+2\mu_{X_{21}}\right),...,{\alpha^2_{MN}}= \exp\left(2x^{(MN)}_{q_{MN}}\sqrt{2\sigma^2_{X_{MN}}}+2\mu_{X_{MN}}\right)\Big).
\end{align}
\hrulefill
\begin{align}\label{app_a_new>>eq3}
& P^{\rm MIMO}_{be,{\rm OC}|b_0,b_k}=\int_{\vec{\boldsymbol{\alpha}}}P^{\rm MIMO}_{be,{\rm OC}|b_0,{\vec{\boldsymbol{\alpha}}},b_k}f_{\vec{\boldsymbol{\alpha}}}(\vec{\boldsymbol{\alpha}})d\vec{\boldsymbol{\alpha}}\approx\frac{1}{{\pi}^{M\times N/2}}\sum_{q_{11}=1}^{U_{11}}w^{(11)}_{q_{11}}\sum_{q_{21}=1}^{U_{21}}w^{(21)}_{q_{21}}...\sum_{q_{MN}=1}^{U_{MN}}w^{(MN)}_{q_{MN}}\times\nonumber\\
& Q\left(\frac{{\underset{j=1}{\overset{N}{\sum}}}{\underset{~i'=1}{\overset{M}{\sum}}}\gamma^{(s)}_{i'\!,j}\exp\left(2x^{(i'\!j)}_{q_{i'\!j}}\sqrt{2\sigma^2_{X_{i'\!j}}}+2\mu_{X_{i'\!j}}\right){\underset{i=1}{\overset{M}{\sum}}}\left(\gamma^{(s)}_{i,j}\!+\!(-1)^{b_0+1}\!\!{\underset{~k=-L_{ij}}{\overset{-1}{\sum}}}\!2b_k\gamma^{(I,k)}_{i,j}\!\right)\!\exp\!\left(\!2x^{(ij)}_{q_{ij}}\!\sqrt{2\sigma^2_{X_{ij}}}\!+\!2\mu_{X_{ij}}\!\right)}{2\sigma_{T_b}\sqrt{\sum_{j=1}^{N}\left(\sum_{i=1}^{M}\gamma^{(s)}_{i,j}\exp\left(2x^{(ij)}_{q_{ij}}\sqrt{2\sigma^2_{X_{ij}}}+2\mu_{X_{ij}}\right)\right)^2}}\!\right).
\end{align}
  \hrulefill
\end{figure*}
Therefore, the ($M\times N$)-dimensional integral of $\int_{\vec{\boldsymbol{\alpha}}}P^{\rm MIMO}_{be|b_0,{\vec{\boldsymbol{\alpha}}},b_k}f_{\vec{\boldsymbol{\alpha}}}(\vec{\boldsymbol{\alpha}})d\vec{\boldsymbol{\alpha}}$ can be calculated as Eq. \eqref{app_a_new>>eq3}.
\section{MGF of the Receiver Output in SISO Scheme}
In this appendix, we calculate the receiver output MGF in SISO scheme. Based on \eqref{r^{b_0}}, conditioned on $\left \{ b_k \right \}_{k=-L}^{-1}$ and $\alpha$, $u^{(b_0)}_{\rm SISO}$ is the sum of two independent RVs. Therefore its MGF ${\Psi }_{u^{(b_0)}_{\rm SISO}}(s)$ is the product of their MGFs, i.e., ${\Psi }_{u^{(b_0)}_{\rm SISO}}(s)={\Psi }_{y^{(b_0)}_{\rm SISO}}\left(s\right)\times {\Psi }_{v_{th}}(s)$.
   We first obtain the conditional MGF of $y^{(b_0)}_{\rm SISO}$ conditioned on $\left \{ b_k \right \}_{k=-L}^{-1}$ and $\alpha$. Then averaging over $\left \{ b_k \right \}_{k=-L}^{-1}$ results the MGF of $y^{(b_0)}_{\rm SISO}$ conditioned on $\alpha $ \cite{einarsson1996principles} as Eq. \eqref{ap_a1}, shown at the top of the next page page.
   \begin{figure*}[!t]
   \normalsize
   \setcounter{equation}{45}
   \begin{align}\label{ap_a1}
     {\Psi }_{{y^{(b_0)}_{\rm SISO}}|\alpha }\left(s\right)&=\E_{b_k}\left[e^{m^{(b_0)}_{\rm SISO}(e^s-1)}|\alpha \right]= \E_{b_k}\left[{\rm exp}\left(\left \{m_{\rm SISO}^{(bd)}+b_0{\alpha}^2m^{(s)}+\sum_{k=-L}^{-1}b_k{\alpha}^2m^{(I,k)} \right \}\left(e^s-1\right)\right){\big|}\alpha\right]\nonumber\\
   &= \exp\left({\left(m_{\rm SISO}^{(bd)}+b_0{\alpha}^2m^{(s)}\right)\left(e^s-1\right)}\right)\times\E_{b_k}\left[\prod_{k=-L}^{-1}{\rm exp}\left( b_k{\alpha}^2m^{(I,k)}\left(e^s-1\right)\right){\big|}\alpha\right].
    \end{align} 
     \hrulefill
   \end{figure*}
Note that $b_k$s are independent Bernoulli RVs with identical probability, i.e., $P_{b_k}(b_k)=\frac{1}{2}\delta(b_k)+\frac{1}{2}\delta(b_k-1)$. Therefore, the latter expectation in \eqref{ap_a1} simplifies to;
\begin{align} \label{ap_a3}
  &\E_{b_k} \Big[\prod_{k=-L}^{-1}{\rm exp} \left( b_k{\alpha}^2m^{(I,k)}\left(e^s-1\right)\right){\big|}\alpha\Big]\nonumber\\&~~~~~~=\prod_{k=-L}^{-1}\E_{b_k}\Big[{\rm exp}\left( b_k{\alpha}^2m^{(I,k)}\left(e^s-1\right)\right){\big|}\alpha\Big]\nonumber\\
   &~~~~~~=\prod_{k=-L}^{-1}(1/2)\Big[1+{\rm exp}\left({\alpha}^2 m^{(I,k)}\left(e^s-1\right)\right)\Big].
 \end{align}
   Finally, inserting \eqref{ap_a3} in \eqref{ap_a1} and then multiplying the result by ${\Psi }_{v_{th}}(s)=\exp({s^2{\sigma }^2_{th}}/{2})$ yields the output MGF as in Eq. \eqref{SISO_MGF}.

\section{MGF of the Receiver Output in MIMO Scheme}
In this appendix, we calculate the MGF of the receiver output in MIMO scheme. Since EGC is used, the combined output of the receiver is $u^{(b_0)}_{\rm MIMO}=\sum^N_{j=1}{u^{(b_0)}_{j}}$. Conditioned on fading coefficients vector $\vec{\boldsymbol{\alpha}}$, received signals from different branches are independent and hence MGF of their sum is the product of each branch's MGF, i.e., ${{\Psi }^{{\rm{EGC}}}_{u^{(b_0)}_{\rm MIMO}|\vec{\boldsymbol{\alpha}}}\left(s\right)}=\prod^N_{j=1}{{\Psi }_{u^{(b_0)}_j|\{\alpha_{ij}\}_{i=1}^{M}}(s)}$. Therefore, we first need to obtain the conditional MGF of each branch. By pursuing similar steps as in Appendix C, $\Psi _{u^{(b_0)}_j|\{\alpha_{ij}\}_{i=1}^{M}}(s)$ can be calculated as Eq. \eqref{ap_b1}, shown at the top of the next page,
\begin{figure*}[!t]
\normalsize
\setcounter{equation}{47}
\begin{align} \label{ap_b1}
  &{\Psi }_{{u^{(b_0)}_j}|\{\alpha_{ij}\}_{i=1}^{M}}\left(s\right)= \exp\left({\frac{{{\sigma}^2_{th}}}{2}s^2}\right)\times
   \E_{b_k}\Bigg[{\rm exp} \Bigg(\Bigg \{ {m_j^{(bd)}}+\sum_{i=1}^{M}\left[b_0{\alpha^2_{ij}}{m_{i,j}^{(s)}}+\sum_{k=-L_{ij}}^{-1}b_k{\alpha^2_{ij}}{m_{i,j}^{(I,k)}}\right] \Bigg \}\left(e^s-1\right)\Bigg)|\{\alpha_{ij}\}_{i=1}^{M}\Bigg]\nonumber\\
   &={\rm exp}\left({\frac{{{\sigma}^2_{th}}}{2}s^2}+\left[{m_j^{(bd)}}+\sum_{i=1}^{M}b_0{\alpha^2_{ij}}{m_{i,j}^{(s)}}\right]\left(e^s-1\right)\right)\times \E_{b_k}\left[\prod_{i=1}^{M}\prod_{k=-L_{ij}}^{-1}{\rm exp}\left(b_k{\alpha^2_{ij}}m_{i,j}^{(I,k)}\left(e^s-1\right)\right){\big|}\{\alpha_{ij}\}_{i=1}^{M}\right],
  \end{align} 
  \hrulefill
\end{figure*}
 where ${m_j^{(bd)}}=({n_b}_j+{n_d}_j)T_b$.
   Supposing the same channel memory as ${L_{ij}}=L_{\rm {max}}$ for all links, performing the latter expectation in \eqref{ap_b1} results the $j$th receiver output MGF as;
\begin{align} \label{ap_b2}
&{\Psi }_{{u^{(b_0)}_j}|\{\alpha_{ij}\}_{i=1}^{M}}\!(s)\!\!=\!{\rm exp}\!\left(\!{\frac{{{\sigma}^2_{th}}s^2}{2}}\!+\!\!\left[\!{m_j^{(bd)}}\!\!+\!\!\sum_{i=1}^{M}\!b_0{\alpha^2_{ij}}{m_{i,j}^{(s)}}\!\right]\!\left(e^s\!\!-\!\!1\right)\!\right)\nonumber\\
  & \times \prod_{k=-L_{\max}}^{-1} (1/2)\left[1+\prod_{i=1}^{M}{\rm exp}\left({\alpha^2_{ij}}m_{i,j}^{(I,k)}\left(e^s-1\right)\right)\right].
\end{align}
{Note that if some of the links have smaller memory, we can add adequate zero components to each of them in order to make all the links with an equal memory length. In other words, if $L_{ij}<L_{\max}$, then $\{m_{i,j}^{(I,k)}\}_{k=-L_{\max}}^{-L_{ij}-1}=0$.} Finally, MGF of the receiver output in MIMO scheme can be obtained as in Eq. \eqref{MGF_MIMO}.

\begin{IEEEbiography}[{\includegraphics[width=1in,height=1.25in,clip,keepaspectratio]{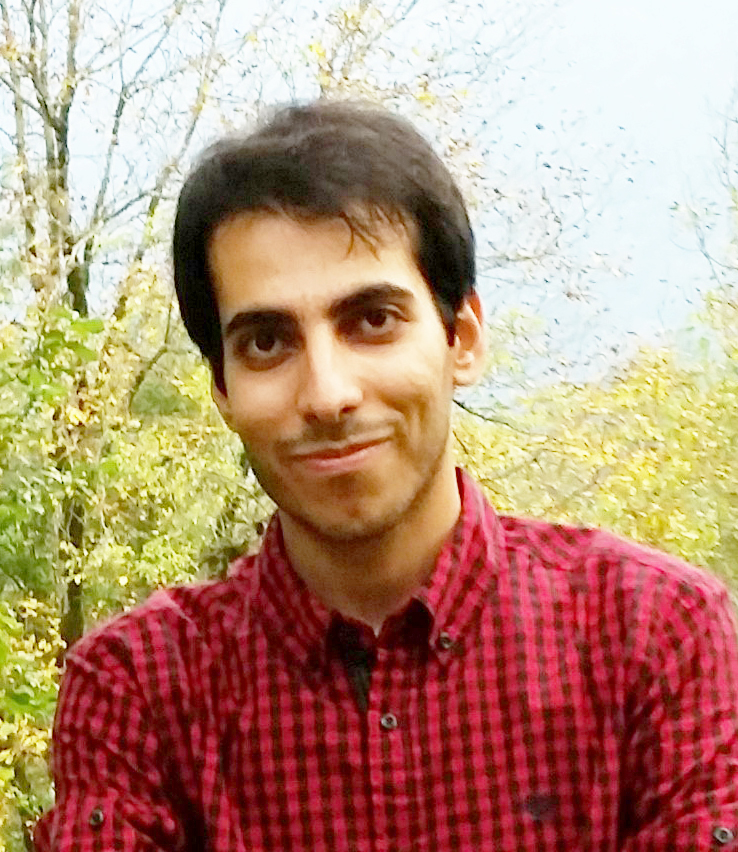}}]{Mohammad Vahid Jamali}
was born in
Talesh, Iran, on May 22, 1991. He received
the B.Sc. degree from K.N. Toosi University of Technology,
Tehran, Iran, in 2013, and the M.Sc. degree from
Sharif University of Technology (SUT), Tehran, Iran, in 2015, both with honors and in Electrical
Engineering.
Since 2013, he has been a member of
the technical staff of the Optical Networks Research
Lab (ONRL) at SUT. 
His general
research interests are in communications theory and
optics with emphasis on wireless applications. Specific research areas include underwater wireless optical communications, mode-division multiplexing in optical fibers, free-space optics, visible light communications, simultaneous wireless information and power transfer, metamaterials and metasurfaces, and DNA sequencing based on optical computing.
\end{IEEEbiography}
\begin{IEEEbiography}[{\includegraphics[width=1in,height=1.25in,clip,keepaspectratio]{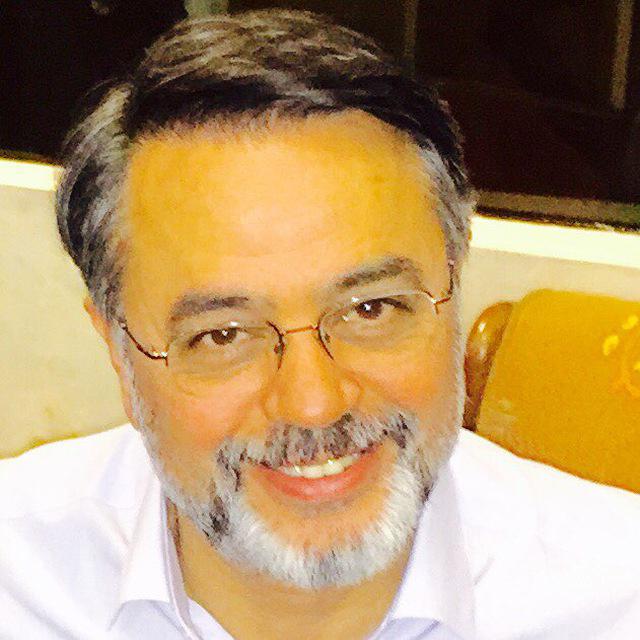}}]{Jawad A. Salehi}
(M'84-SM'07-FM'10) was born in
Kazemain, Iraq, on December 22, 1956. He received
the B.Sc. degree from the University of California,
Irvine, in 1979, and the M.Sc. and Ph.D. degrees from
the University of Southern California (USC), Los
Angeles, in 1980 and 1984, respectively, all in electrical
engineering. He is currently a Full Professor at
the Optical Networks Research Laboratory (ONRL),
Department of Electrical Engineering, Sharif University
of Technology (SUT), Tehran, Iran. From 1981
to 1984, he was a Full-Time Research Assistant at
the Communication Science Institute, USC. From 1984 to 1993, he was a
Member of Technical Staff of the Applied Research Area, Bell Communications
Research (Bellcore), Morristown, NJ. Prof. Salehi was an Associate
Editor for Optical CDMA of the IEEE Transactions on Communications,
since 2001 to 2012. He is among the 250 preeminent and most
influential researchers worldwide in the Institute for Scientific Information (ISI)
Highly Cited in the Computer-Science Category.
\end{IEEEbiography}
\begin{IEEEbiography}[{\includegraphics[width=1in,height=1.25in,clip,keepaspectratio]{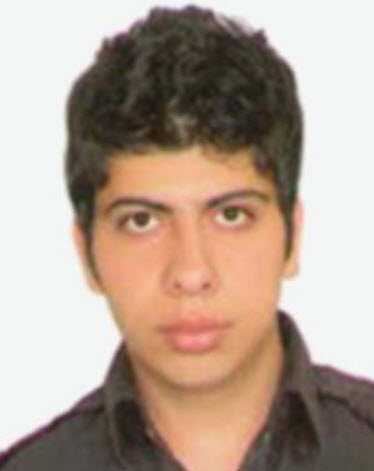}}]{Farhad Akhoundi} received the B.Sc. degree (first-class honor) from Shahid Rajaee Teacher Training University (SRTTU), Tehran, in 2010, and the M.Sc. degree from Sharif University of Technology (SUT), Tehran, Iran, in 2012, both in electrical engineering. From 2012 to 2014, he was a member of the technical staff of the Optical Networks Research Laboratory (ONRL) at SUT. From 2014 to 2015, he was employed by DNA Microarray Analysis Lab (DMA Lab), a knowledge-based company in SUT, working on camera-based DNA microarray scanner. Farhad is currently working toward the Ph.D. degree in the University of Arizona (UA), Tucson, USA. He is also a Full-Time Graduate Research Assistant at College of Optical Sciences, UA. His research interests are in the areas of fiber optics, visible light, and underwater wireless optical multiple access networks. He is also interested in multi-photon microscopy and fluorescence imaging in the biosciences.
\end{IEEEbiography}

\end{document}